\documentclass[aps,prl,twocolumn,showpacs,amsmath,amssymb,superscriptaddress]{revtex4-1}

\usepackage{graphicx}
\usepackage{color}
\usepackage{epstopdf}
\usepackage{listings}

\newcommand{\refeq}[1]{Eq.~(\ref{#1})}

\begin{document}

\title{Self-organization of light in optical media with competing nonlinearities}
\author{F. Maucher}
\affiliation{Joint Quantum Centre (JQC) Durham-Newcastle, Department of Physics, Durham University, Durham DH1 3LE, United Kingdom}
\affiliation{Department of Mathematical Sciences, Durham University, Durham DH1 3LE, United Kingdom}
\author{T. Pohl}
\affiliation{Max Planck Institute for the Physics of Complex Systems, 01187 Dresden,
Germany}
\author{S. Skupin}
\affiliation{Univ.~Bordeaux - CNRS - CEA, Centre Lasers Intenses et Applications, UMR 5107, 33405 Talence, France}
\author{W. Krolikowski}
\affiliation{Laser Physics Centre, Research School of
Physics and Engineering, Australian National University, Canberra, ACT
0200, Australia}
\affiliation{Science Program, Texas A\&M University at Qatar, Doha, Qatar}

\begin{abstract}
We study the propagation of light beams through optical media with competing nonlocal nonlinearities. 
We demonstrate that the nonlocality of competing focusing and defocusing nonlinearities gives rise to self-organization and stationary states with stable hexagonal intensity patterns, 
akin to transverse crystals of light filaments. Signatures of this long-range ordering are shown to be observable in the propagation of light in optical waveguides and even in free space. We consider a specific form of the nonlinear response that arises in atomic vapor upon proper light coupling. Yet, the general phenomenon of self-organization is a generic consequence of competing nonlocal nonlinearities, and may, hence, also be observed in other settings. 
\end{abstract}
\maketitle

Self-organization constitutes one of the most fascinating phenomena appearing in nonlinear systems. 
During the process, strong interactions among the system components lead to the formation of spatial structures and long-range ordering.
This effect plays a crucial role in a broad context, from biology~\cite{Self_organization_biology,Rietker:science:04,Escaff:PRE:2015},
chemistry~\cite{Meron:pr:92,Swinney:nature:97} and hydrodynamics~\cite{Newel:arfm:93} to soft-matter physics~\cite{Likos:PhysRep:2001,Leunissen:nature:05,Liu:pre:08}.
In optics the spontaneous formation of regular intensity patterns has been observed almost 30 years ago~\cite{Grynberg:oc:88}, and since been explored in various  
settings~\cite{Ulf:pre:97,Arecchi:pr:99,Lodahl:pra:1999,Camara:PRA:2015}. 
Common to all these experiments is the requirement of an appropriate feedback mechanism, 
provided e.g. by an optical cavity or a single mirror that retro-reflects traversing light back into the medium, while feedback-less pattern formation in a Kerr medium has been observed~\cite{Bennink:PRL:2002} from far-field interference of small-scale regular filaments.
On the other hand, the formation of spatial structures solely due to the nonlinear propagation of light has attracted great interest over the past years \cite{Trillo:SS:01,KivsharAgrawal:2003}.
Most prominently, optical solitons emerging from local Kerr-type nonlinearities of various kinds have been actively investigated~\cite{Quiroga:josab:97,Malomed:phys_d:02,Corney:pre:01} and play an important role for intense light propagation \cite{Couairon:pr:441:47} and potential applications to fiber optics communication \cite{Agrawal:NFO:01}.  Nonlinearities can also cause extended structures to emerge, e.g., from modulation instabilities (MI) that drive a growth of broad-band density modulations and ultimately lead to the formation of randomly arranged filaments \cite{Mamaev:PRA:1996,Saffman:JOB:2004,Meier:PRL:2004,Henin:apb:100:77}.

\begin{figure*}
\includegraphics[width=\textwidth]{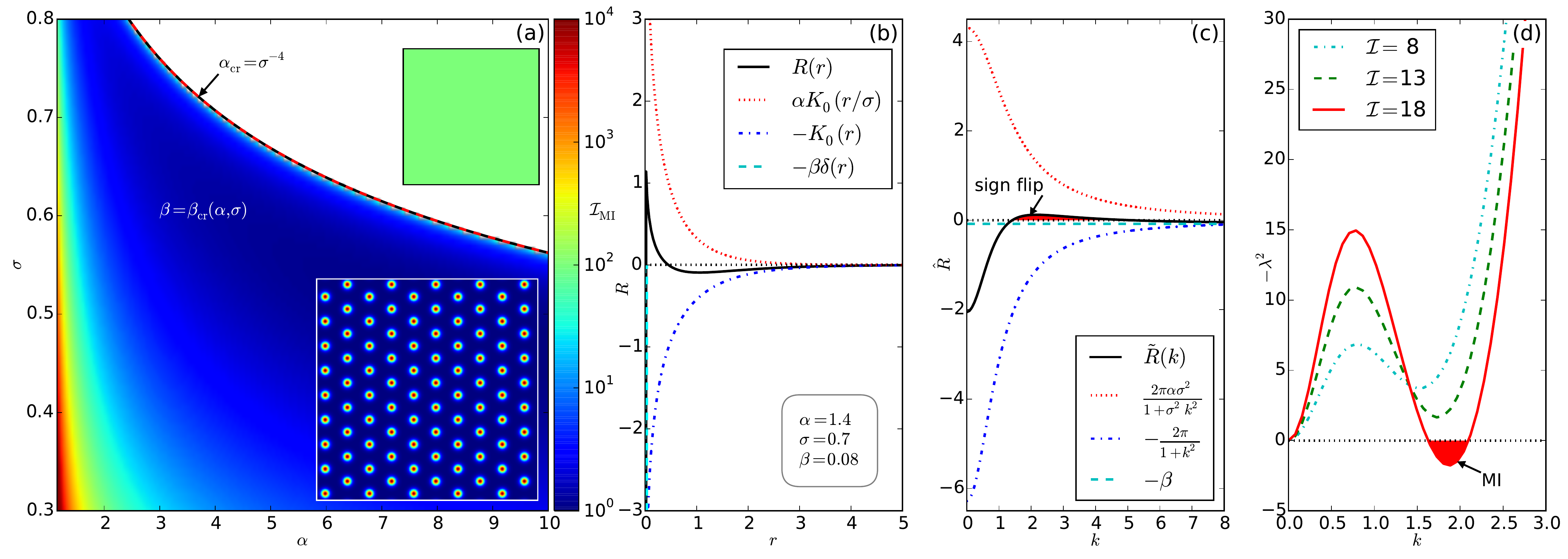}
\caption{(color online) (a) Where MI occurs in a finite momentum range and ordered intensity patterns (lower inset) are possible. The color coding for $\alpha<\alpha_{\rm cr}$ shows the minimum intensity for MI, while plane wave solutions remain stable for $\alpha>\alpha_{\rm cr}$.  (b) Position [Eq.~(\ref{eq:response})] and (c) momentum space [Eq.~(\ref{eq:F_responses})] form of the nonlinear response function (solid line) arising from a combination of a nonlocal focusing (dotted line), nonlocal defocusing (dashed-dotted line) and local defocusing nonlinearity (dashed line). Panel (d) shows the corresponding dispersion relation, Eq.~(\ref{eq:growth_rate}), of periodic perturbations with momentum $k$.
\label{fig:fig0}}
\end{figure*}

In this work, we show that self-organization into spatially ordered patterns [see Fig.~\ref{fig:fig0}(a)] of unidirectionally propagating light can occur in media with a spatially nonlocal nonlinearity. Although the absence of any feedback mechanism in our system may be expected to prevent the formation of extended patterns \cite{OptRes}, we show that this is not the case and regular patterns can arise from a suitably designed nonlocality of the medium. This sets it apart from previously studied systems~\cite{Grynberg:oc:88,Ulf:pre:97,Arecchi:pr:99,Lodahl:pra:1999,Camara:PRA:2015}, and as we will see below, implies profound changes of the underlying physics, including the threshold behaviour for optical pattern formation \cite{ISI:A1988R274000014, ISI:A1994PL62700090, ISI:A1992GY19100011}.
The effect rests upon a sign change of the optical response in Fourier space~\cite{Likos07}, which in the present case drives MI within a finite band of momenta [see Fig.~\ref{fig:fig0}(c,d)]. 
This condition provides a challenge for most nonlinear optics experiments where nonlocality typically arises from 
transport processes \cite{Ultanir:ol:03,Dabby:apl:13:284,Derrien:JOpt:2000,Suter:1993,Skupin:prl:2007,Conti:PRL:2005,Conti:PRL:2006b} 
that naturally yield a sign-definite nonlinear response. Overcoming this obstacle, we consider a combination of a focusing and defocusing nonlinearity [see Fig.~\ref{fig:fig0}(b)] and describe a physical realization of the proposed response function in atomic vapour. We derive simple conditions for the emergence of stable ordered states and show that signatures of such ''crystals'' are observable in the propagation of light through the medium.

Specifically, we study the evolution of a wave function $\psi({\bf r},z)$, representing the slowly varying envelope of the electric field component of a light beam. Its propagation is governed by the nonlinear Schr\"odinger equation 
\begin{equation}\label{eq:nls}
\begin{split}
i\partial_z \psi({\bf r},z) & =-\Delta_\perp \psi({\bf r},z)+ U({\bf r})\psi({\bf r},z) - \frac{i}{\ell}\psi({\bf r},z)\\
& \quad -\int R(|{\bf r-r^\prime}|) |\psi({\bf r}^\prime,z)|^2 d^2r^\prime \psi({\bf r},z),
\end{split}
\end{equation}
with ${\bf r}$ and $z$ denoting generalized transverse and longitudinal (propagation) coordinates, respectively. The amplitude $\psi({\bf r},z)$ and all other parameters in Eq.~(\ref{eq:nls}) represent dimensionless quantities as obtained from proper length- and time-scaling of the specific realization given in the Supplement~\cite{suppl}.
The parameter $\ell$ is the linear absorption length and the external potential $U({\bf r})$ may represent an additional optical waveguide.
We consider a response function of the cubic nonlinearity
\begin{equation}
R(r)=\alpha K_0\left( \frac{r}{\sigma} \right) - K_0\left( r \right)-\beta\delta(r),\label{eq:response}
\end{equation}
that is composed of three terms. The first and the second term describe a focusing and defocusing nonlocal nonlinearity, respectively, while the third corresponds to a local defocusing nonlinearity as given by the Dirac delta function, $\delta(r)$. The parameter $\beta>0$ represents its strength and $K_0$ denotes the modified Bessel function of the second kind. Scaling with respect to the defocusing nonlinearity leaves two parameters, describing the strength ($\alpha>0$) and spatial range ($0<\sigma<1$) of the focusing nonlinear response relative to that of the defocusing term \cite{suppl}.
While our general findings do not depend qualitatively on the shape of the  nonlocal kernel, the function $K_0(r)$ plays an important role in diverse optical settings. For example, it describes light propagation in nematic liquid crystals with orientational nonlinear response~\cite{Conti:prl:2003}, and was used to model the nonlinearity of thermal media~\cite{Ghofraniha:PRL:2007,Conti:prl:09}. 
Although most of these situations only yield a single sign-definite response, a combination of both appears possible~\cite{Warenghem:mclc:06,Warenghem:josab:08}.

Here, we suggest that the complete response function,  \refeq{eq:response}, can be realized in alkali metal vapor. One can obtain a  cubic Kerr nonlinearity whose nonlocal character emerges from diffusive atomic motion. In fact, the formation of nonlocal solitons due to a response function $\sim K_0(r)$ in such systems has already been demonstrated experimentally \cite{Suter:1993}. As we show in~\cite{suppl}, the simultaneous coupling of light to near-resonant transitions involving two incoherently coupled hyperfine levels can give rise to competing nonlinearities as given in Eq.~(\ref{eq:response}). Choosing the frequency detuning of the propagating light just in between the corresponding hyperfine splitting yields a blue and red detuned transition and, thereby, two nonlocal nonlinearities of opposite sign. Moreover, the devised approach naturally provides a third, local nonlinearity, which plays a critical role for the emergence and stability of regular patterns, as we discuss below.

To this end, it appears appropriate to first consider $U({\bf r})=\ell^{-1}=0$. The aforementioned MI refers to linear instability of plane wave solutions $\psi_{\rm pw}({\bf r},z)=A_0\exp \left(\mathrm{i} \mu z \right)$ with respect to periodic modulations $a({\bf r},z) = a_1\exp(i{\bf k r} + \lambda z)+a_2^*\exp(-i{\bf k r} + \lambda^* z)$~\cite{Benjamin:jfm:67}, where $\mu=A_0^2\int R(|{\bf r}|)d^2r$ is the propagation constant. Linearization in terms of the perturbation amplitudes $a_{1,2}$ then yields the growth rate, $\lambda$, 
\begin{equation}
 \lambda^2=-k^2\left( k^2 -2\mathcal{I}\tilde{R}(k) \right),
\label{eq:growth_rate}
\end{equation}
of a given mode with wave vector ${\bf k}$, and $\mathcal{I}=|A_0|^2$ is the plane wave intensity. The Fourier transform, $\tilde{R}({\bf k})$, of the response function Eq.~(\ref{eq:response}) reads
\begin{equation}
\tilde{R}(k)=
\frac{2\pi\alpha\sigma^2}{1+\sigma^2k^2}-\frac{2\pi}{1+k^2}-\beta.\label{eq:F_responses}
\end{equation}
Wherever $\tilde{R}(k)>0$, one can find MI, i.e. a real and positive growth rate $\lambda$, for a sufficiently large intensity, $\mathcal{I}$, of the initial plane wave solution. In particular, if $\tilde{R}(0)<0$ and $\tilde{R}(k)$ changes sign at a finite value of $k=k_0>0$, MI only occurs in a {\em finite} band of wavelengths $<2\pi/k_0$~\cite{esbensen:pra:11}. Fig.~\ref{fig:fig0}(d) shows a typical spectrum and illustrates the onset of MI as $\mathcal{I}$ is increased above the critical intensity $\mathcal{I}_{\rm MI}$. The resulting wavenumber filtering is important as it yields an additional length scale emerging from initial white-noise perturbations which are typically present in experiments. On the contrary, more common long-wavelength MI requires overall focusing nonlinearities [$\tilde{R}(0)>0$] and includes arbitrarily small wavenumbers in the instability interval. This results in an infinite band of unstable wavelengths, associated with random filamentation and, ultimately, the formation of bright solitons or collapse~\cite{Bespalov:JETP:1966,Berge:pr:303:259}. 

For our choice of response function, $-\lambda^2(k)$ exhibits a local maximum followed by a minimum [Fig.~\ref{fig:fig0}(d)], which bears analogies to the known maxon-roton structure of excitation spectra known for superfluid Helium \cite{Landau41,Feynman56} and studied for Bose-Einstein condensates with finite-range interactions \cite{Santos:prl:03,Henkel:PRL:2010}. 
The roton minimum and the associated instability in quantum fluids may appear as a precursor to a solid phase \cite{Schneider71,Astrakharchik07}, but can also usher in a transition to a modulated fluid described by a single-particle amplitude $\psi$ \cite{Henkel:PRL:2010,Cinti14}.

\begin{figure}
\includegraphics[width=\columnwidth]{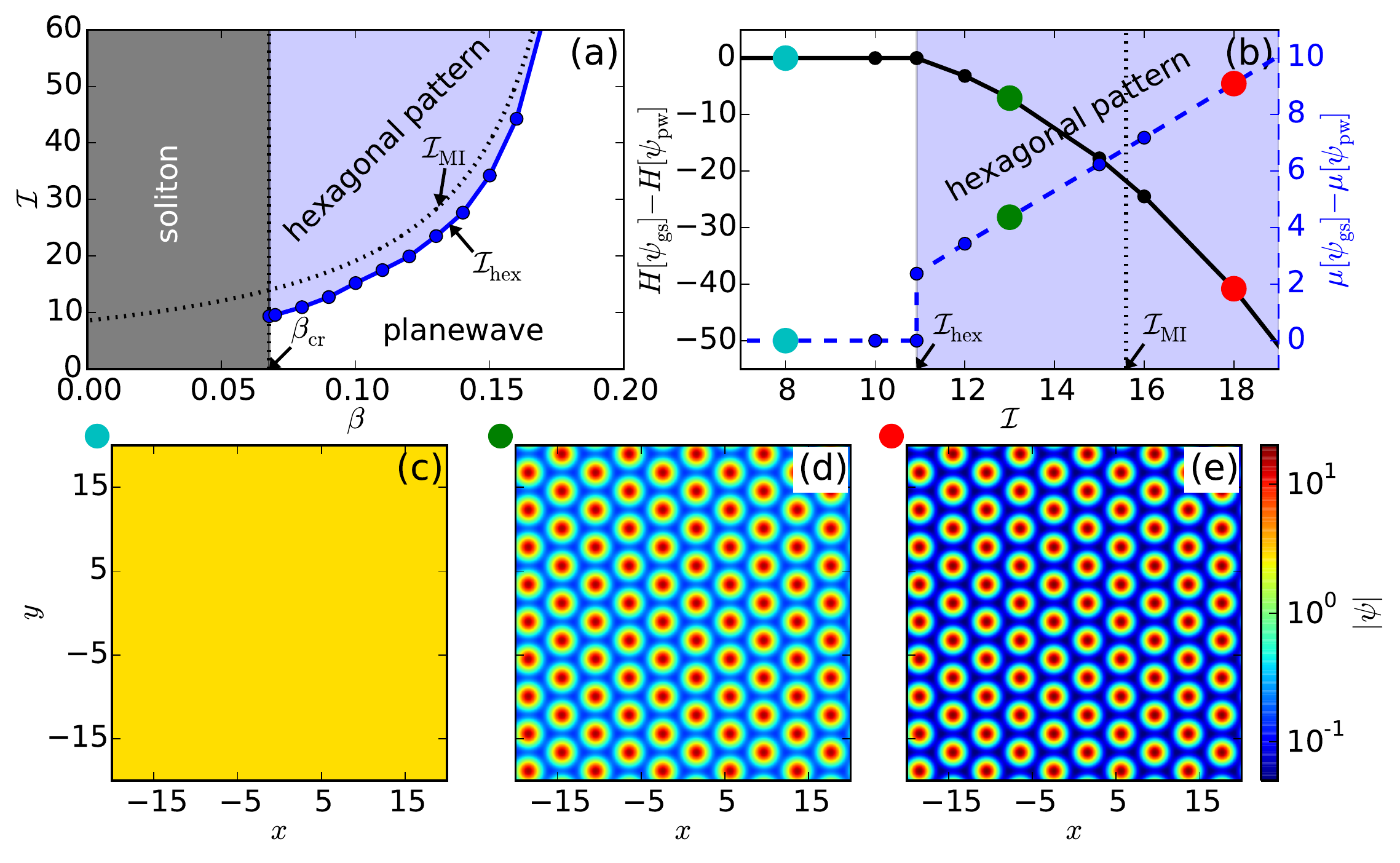}
\caption{(color online) (a) Phase diagram for $\alpha=1.4$, $\sigma=0.7$ illustrating the emergence of three different phases from the plane wave solution as a function of $\mathcal{I}$ and $\beta$. (b) Difference in Hamiltonian density $H$ (solid line) and propagation constant $\mu$ (dashed line) between plane wave solutions $\psi_{\rm pw}$ and numerically computed ground states $\psi_{\rm g}$ versus plane wave intensity $\mathcal{I}$ ($\beta=0.08$). 
Pattern formation at the threshold intensity  $\mathcal{I}_{\rm hex}$ is accompanied by a jump in the propagation constant,
and occurs well below the critical intensity for MI. Exemplary ground states for different plane wave intensities
$\mathcal{I}$ obtained from imaginary propagation (see text) are shown in (c-e). 
\label{fig:fig1}}
\end{figure}

In order to further analyze the present system, we consider the ground state of Eq.~(\ref{eq:nls}), i.e. the minimizer of the Hamiltonian density 
\begin{equation}\label{eq:hamiltonian}
\begin{split}
H&=\frac{1}{V} \int \left|\nabla\psi_{\rm st}({\bf r})\right|^{2}{\rm d}^2r \\
& \quad -\frac{1}{2V}
\iint  R(\mathbf{r}-\mathbf{r}^\prime)\left|\psi_{\rm st}({\bf r})\right|^{2}|\psi_{\rm st}(\mathbf{r}^\prime)|^{2}{\rm d}^2r^\prime {\rm d}^2r
\end{split}
\end{equation}
in the limit of a large integration area $V\rightarrow\infty$. Since we are looking for a stationary solution, $ \psi_{\rm st}=A_{\rm st}({\bf r})e^{i\mu z}$, the Hamiltonian is only affected by the transverse profile $A_{\rm st}({\bf r})$. 
The analysis of Eq.~(\ref{eq:hamiltonian}) reveals a rich ground state behavior, including plane waves, hexagonal intensities patterns as well as bright soliton solutions. Figure~\ref{fig:fig1}(a) illustrates the emergence of these different phases from the plane wave solution as a function of the plane wave intensity $\mathcal{I}$ and the strength $\beta$ of the local defocusing nonlinearity. For $0<\sigma<1$ and $\alpha>1$, the nonlocal part of the kernel Eq.~(\ref{eq:response}) diverges to positive values as $r\rightarrow0$, which inevitably leads to the existence of a bright soliton as groundstate under the sole action of the nonlocal nonlinearity. Fortunately, the additional local nonlinearity $\sim\beta$ tends to diminish this short-distance focusing behavior and ultimately allows to suppress the soliton solution upon exceeding a critical local defocusing $\beta_{\rm cr}$. We can estimate this critical value from below through a variational analysis of the minimizer of 
 Eq.~(\ref{eq:hamiltonian}), assuming a Gaussian form of $A_{\rm st}({\bf r})$ (see \cite{suppl} for further details). This calculation typically yields a good estimate of the exact $\beta_{\rm cr}$ obtained from numerical simulations, e.g., $\beta_{\rm cr}^{\rm (var)}\approx0.0654$ and $\beta_{\rm cr}^{\rm (num)}\approx0.0678$ in Fig.~\ref{fig:fig1}(a). 

Having obtained $\beta_{\rm cr}$ as a function of $\alpha$ and $\sigma$ we can calculate the critical intensity $\mathcal{I}_{\rm MI}$ necessary to induce finite-$k$ MI at the minimum value of $\beta=\beta_{\rm cr}$. The result, shown in Fig.~\ref{fig:fig0}(a), indeed yields an extended range of parameters where a modulated ground state is possible without contracting to a single bright soliton. 
We find that the transition line which separates MI from the region where an initial plane wave will remain stable for every value of $\mathcal{I}$ follows a simple relation which
can be derived from the following argument. Noting that the nonlocal response asymptotically decreases as $2\pi(\alpha-1)/k^2>0$ it needs to exhibit a local minimum at $k=0$ in order to allow for a finite-$k$ sign change through the addition of the local defocusing nonlinearity. 
Formally, this requirement corresponds to $\partial^2_k \tilde{R}(k)|_{k=0}>0$ and, thus, yields $\alpha_{\rm cr}=\sigma^{-4}$. Alternatively, we can determine the transition line by excluding the possibility of long-wavelength MI which implies $\tilde{R}(0)\le0$. Since both criteria are equivalent, their combination yields the critical $\beta_{cr}=2\pi(\sigma^{-2}-1)$ along the transition line. This expression matches our numerical results and coincides with the variational analysis described above (see \cite{suppl}).

To determine the ground state $\psi_{\rm gs}$ we solve Eq.~(\ref{eq:nls}) for an imaginary propagation coordinate ($z\rightarrow-iz$) with periodic boundary conditions and $U({\bf r})=\ell^{-1}=0$, starting from a plane wave, $\psi({\bf r},0)=\mathcal{I}^{1/2}+\varepsilon({\bf r})$, perturbed by small amplitude white noise, $\varepsilon({\bf r})$. Above the threshold intensity  $\mathcal{I}_{\rm hex}<\mathcal{I}$ we find that the ground state $\psi_{\rm gs}$ acquires  hexagonal intensity pattern as shown in Figs.~\ref{fig:fig1}(d,e). This threshold value $\mathcal{I}_{\rm hex}$
is significantly smaller than the critical intensity for MI. While the plane wave solution remains stable for $\mathcal{I}_{\rm hex}<\mathcal{I}<\mathcal{I}_{\rm MI}$, it, consequently, ceases to be the lowest-energy state in this intensity region. 
We can detect the ground state transition by monitoring Hamiltonian density $H[\psi_{\rm gs}]$ and propagation constant $\mu[\psi_{\rm gs}]$ relative to those of the plane wave solution $\psi_{\rm pw}$. The found behavior, shown in Fig.~\ref{fig:fig1}(b) is consistent with a first order phase transition as expected for two-dimensional systems \cite{Macri13,Cinti14b}. As a result, intensity modulations in $\psi_{\rm gs}$ set in abruptly upon crossing $\mathcal{I}_{\rm hex}$ rather then growing continuously. 

While MI, hence, represents a sufficient, but not necessary criterion for structured ground states, the phase transition occurs as a precursor of the instability and does not take place in systems which do not feature finite-$k$ MI. We also note that the intensity patterns can neither be interpreted in terms of conventional bright solitons, nor do they represent dark solitons since the found state does not feature any phase structure which is typical for the latter. These observations underline again the importance of the competition between the nonlocal nonlinearities to observe the described phenomena.

\begin{figure}
\includegraphics[width=\columnwidth]{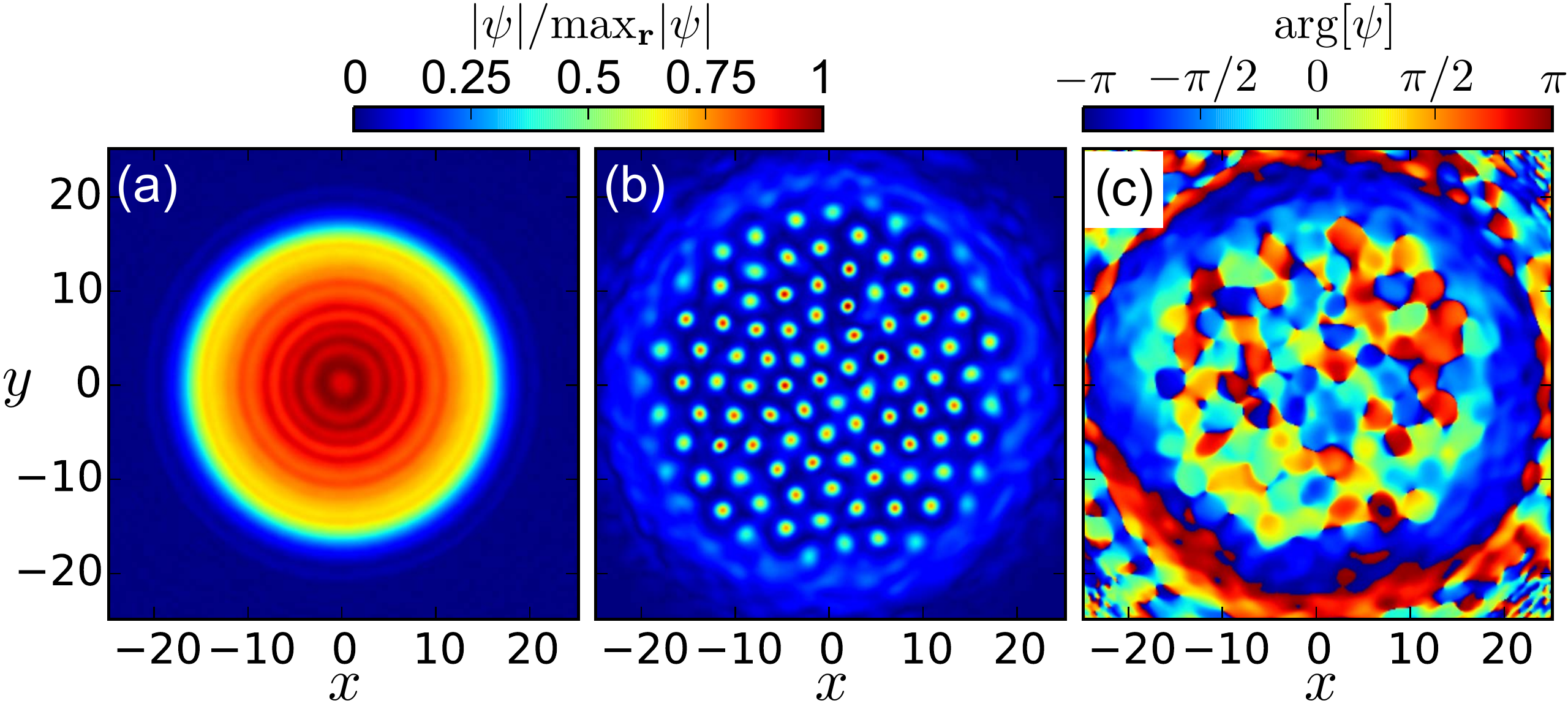}
\caption{(color online) Guided light propagation for $\alpha=1.4$, $\sigma=0.7$, $\beta=0.08$, $U({\bf r})=(r/4)^2$, $\ell^{-1}=0$ and two different intensities of (a) $\mathcal{I}=10<\mathcal{I}_{\rm MI}$ and (b,c) $\mathcal{I}=20>\mathcal{I}_{\rm MI}$ after a propagation length of $z=10$. 
Panel (c) indicates the inhomogeneous phase evolution accompanying the emergence of hexagonal intensity patterns shown in (b). (a) Below $\mathcal{I}_{\rm MI}$ the intensity profile develops a weak ring structure due to the initial noise. See \cite{suppl} for further details.
\label{fig:fig2}}
\end{figure}

Let us now study signatures of these stationary properties in the propagation of light, that would potentially be observable in experiments. We begin with the real space propagation of Eq.~(\ref{eq:nls}) in a hollow-core optical waveguide, which we model by a simple harmonic potential $U({\bf r})=(r/4)^2$. As the initial condition, we choose a Thomas-Fermi profile $\psi({\bf r},0) = \mathcal{I}^{1/2}\sqrt{1-\frac{r^2}{w^2}}+\varepsilon({\bf r})$, whose width $w=4\sqrt{-\mathcal{I}\tilde{R}(0)}$ is determined by the confining potential, the intensity $\mathcal{I}$, and $\tilde{R}(0)=\int R(r){\rm d}^2r<0$. Figure~\ref{fig:fig2} shows  intensity profiles obtained for different input intensities $\mathcal{I}$ below and above $\mathcal{I}_{{\rm MI}}$. While the former case preserved the rotational symmetry and yields a nearly stationary intensity profile [Fig.~\ref{fig:fig2}(a)], the higher intensity results in the formation of regularly spaced filaments [Fig.~\ref{fig:fig2}(b)]. Dynamically, pattern formation is preceded by MI leading to rapid formation of filaments. Due to the nonlocal nonlinearity and its overall defocusing character the formed filaments experience effective repulsive interactions and eventually settle into a hexagonal lattice structure. Note, that we have set $\ell^{-1}=0$ in order to study the dissipationless propagation dynamics. Nevertheless, ordering is still possible since the associated Hamiltonian density is dissipated into phase gradients \cite{suppl} that predominantly emerge in the low-intensity regions between the filaments [Fig.~\ref{fig:fig2}(c)].

\begin{figure}
\includegraphics[width=\columnwidth]{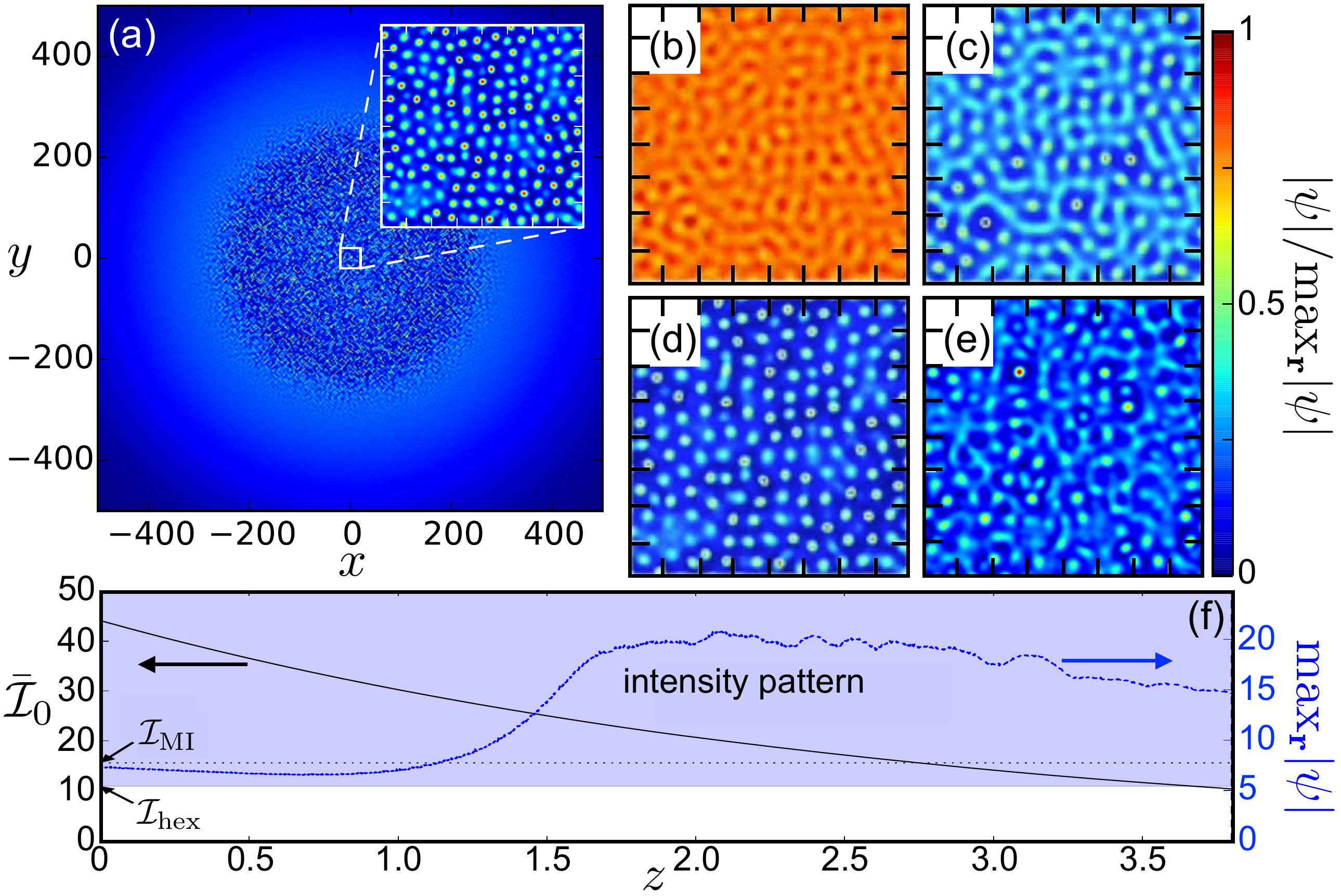}
\caption{(color online) Free propagation for $\alpha=1.4$, $\sigma=0.7$, $\beta=0.08$, $U=0$, $\ell=5.3$. During propagation regular intensity patterns form in the beam center, as shown in the inset of (a). Panels (b-e) show the propagation dynamics in this central region of area $V_0$ for different propagation lengths $z=1$ (b), $1.7$ (c), $2.3$ (d) and $3.7$ (e). (f) Evolution of the peak amplitude, ${\rm max}_{\bf r}|\psi|$, and average intensity, $\bar{\mathcal{I}}_0$, in the central area $V_0$. See \cite{suppl} for further details.
\label{fig:fig3}}
\end{figure}

In Fig.~\ref{fig:fig3}, we show the propagation dynamics for an input beam 
$\psi({\bf r},z=0) = \mathcal{I}^{1/2} \exp\left[-\frac{r^4}{w^4}\right]+\varepsilon({\bf r})$, with $\mathcal{I}=40$ and $w=500$, for $U=0$ and $\ell=5.3$. Again one finds fast filamentation, as indicated by the peak amplitude dynamics shown in Fig.~\ref{fig:fig3}(c). 
Subsequently, the filaments start to form short-range ordered structures. However, this self-organized state cannot be sustained against intensity-loss due to absorption and beam spreading. 
It ultimately disintegrates once the average intensity, $\bar{\mathcal{I}}_0(z)=V_0^{-1}\int_{V_0}|\psi({\bf r},z)|^2{\rm d}^2r$, in the central area, $V_0$, approaches $\mathcal{I}_{\rm hex}$. 

We finally want to relate these findings to the proposed experimental realization in atomic media.  As further detailed in the Supplement~\cite{suppl}, the parameters used in Figs.~\ref{fig:fig2} and \ref{fig:fig3} can be obtained for a Sodium vapor at a density of $9\times10^{13}$~cm$^{-3}$ where incoherent hyperfine pumping with a rate of $2\pi\times0.9$~MHz transfers population from the $|F=1\rangle$ to the $|F=2\rangle$ state, and vice versa with a rate of $2\pi\times3.9$~MHz. Coupling the light field detuned by $2\pi\times14$~MHz from the $D_1$ transition then yields $\alpha=1.4$, $\sigma=0.7$, and a dimensionless absorption length of $\ell=5.3$. The dimensionless intensity $\mathcal{I}=40$ then gives the reasonable value of $300$~W/cm${^2}$. For a diffusion constant of $30$~cm$^2/$s the dimensionless unit length corresponds to $10~\mu$m, making the predicted patterns observable with conventional imaging techniques. Generally, the number of tunable parameters entails considerable flexibility, allowing to find viable experimental conditions for other combinations of $\alpha$ and $\sigma$ as well.

In summary,  we have investigated the emergence of crystalline intensity patterns due to a competition of nonlocal optical nonlinearities with different signs and ranges. The phenomenon was traced back to a first-order phase transition between ground states of the underlying propagation equation. Yet, we showed that it should be observable in the unidirectional propagation of light, facilitated by Hamiltonian density dissipation into phase gradients. We have devised a physical implementation in dilute atomic vapor that realizes the proposed model. However, the presented analysis also applies to other media in which light propagation is adequately described by Eq.(\ref{eq:nls}). We hence expect this work to be relevant to such systems where competing nonlocal nonlinearities may arise from different transport mechanisms, including  particle or heat diffusion or reorientation of induced dipoles.

\begin{acknowledgments}
Numerical simulations were performed using computing
resources at M\'esocentre de Calcul Intensif Aquitain (MCIA) and Grand Equipement National pour le Calcul Intensif (GENCI, grant no.~2015-056129).
This work was partially supported by the Qatar National Research Fund through the National  Priorities  Research  Program.
FM acknowledges funding by the Leverhulme Trust Research Programme Grant RP2013-K-009.
\end{acknowledgments}

\bibliographystyle{apsrev4-1}
\bibliography{thesis}

\begin{thebibliography}{57}%
\makeatletter
\providecommand \@ifxundefined [1]{%
 \@ifx{#1\undefined}
}%
\providecommand \@ifnum [1]{%
 \ifnum #1\expandafter \@firstoftwo
 \else \expandafter \@secondoftwo
 \fi
}%
\providecommand \@ifx [1]{%
 \ifx #1\expandafter \@firstoftwo
 \else \expandafter \@secondoftwo
 \fi
}%
\providecommand \natexlab [1]{#1}%
\providecommand \enquote  [1]{``#1''}%
\providecommand \bibnamefont  [1]{#1}%
\providecommand \bibfnamefont [1]{#1}%
\providecommand \citenamefont [1]{#1}%
\providecommand \href@noop [0]{\@secondoftwo}%
\providecommand \href [0]{\begingroup \@sanitize@url \@href}%
\providecommand \@href[1]{\@@startlink{#1}\@@href}%
\providecommand \@@href[1]{\endgroup#1\@@endlink}%
\providecommand \@sanitize@url [0]{\catcode `\\12\catcode `\$12\catcode
  `\&12\catcode `\#12\catcode `\^12\catcode `\_12\catcode `\%12\relax}%
\providecommand \@@startlink[1]{}%
\providecommand \@@endlink[0]{}%
\providecommand \url  [0]{\begingroup\@sanitize@url \@url }%
\providecommand \@url [1]{\endgroup\@href {#1}{\urlprefix }}%
\providecommand \urlprefix  [0]{URL }%
\providecommand \Eprint [0]{\href }%
\providecommand \doibase [0]{http://dx.doi.org/}%
\providecommand \selectlanguage [0]{\@gobble}%
\providecommand \bibinfo  [0]{\@secondoftwo}%
\providecommand \bibfield  [0]{\@secondoftwo}%
\providecommand \translation [1]{[#1]}%
\providecommand \BibitemOpen [0]{}%
\providecommand \bibitemStop [0]{}%
\providecommand \bibitemNoStop [0]{.\EOS\space}%
\providecommand \EOS [0]{\spacefactor3000\relax}%
\providecommand \BibitemShut  [1]{\csname bibitem#1\endcsname}%
\let\auto@bib@innerbib\@empty
%</preamble>
\bibitem [{\citenamefont {Camazine}\ \emph {et~al.}(2001)\citenamefont
  {Camazine}, \citenamefont {Deneubourg}, \citenamefont {Nigel}, \citenamefont
  {Sneyd}, \citenamefont {Theraulaz},\ and\ \citenamefont
  {Bonabeau}}]{Self_organization_biology}%
  \BibitemOpen
  \bibfield  {author} {\bibinfo {author} {\bibfnamefont {S.}~\bibnamefont
  {Camazine}}, \bibinfo {author} {\bibfnamefont {J.-L.}\ \bibnamefont
  {Deneubourg}}, \bibinfo {author} {\bibfnamefont {R.~F.}\ \bibnamefont
  {Nigel}}, \bibinfo {author} {\bibfnamefont {J.}~\bibnamefont {Sneyd}},
  \bibinfo {author} {\bibfnamefont {G.}~\bibnamefont {Theraulaz}}, \ and\
  \bibinfo {author} {\bibfnamefont {E.}~\bibnamefont {Bonabeau}},\ }\href@noop
  {} {\emph {\bibinfo {title} {Self-organization in biological systems}}}\
  (\bibinfo  {publisher} {Princeton University Press},\ \bibinfo {year}
  {2001})\BibitemShut {NoStop}%
\bibitem [{\citenamefont {Rietkerk}\ \emph {et~al.}(2004)\citenamefont
  {Rietkerk}, \citenamefont {Dekker}, \citenamefont {de~Ruiter},\ and\
  \citenamefont {van~de Koppel}}]{Rietker:science:04}%
  \BibitemOpen
  \bibfield  {author} {\bibinfo {author} {\bibfnamefont {M.}~\bibnamefont
  {Rietkerk}}, \bibinfo {author} {\bibfnamefont {S.~C.}\ \bibnamefont
  {Dekker}}, \bibinfo {author} {\bibfnamefont {P.~C.}\ \bibnamefont
  {de~Ruiter}}, \ and\ \bibinfo {author} {\bibfnamefont {J.}~\bibnamefont
  {van~de Koppel}},\ }\href@noop {} {\bibfield  {journal} {\bibinfo  {journal}
  {Science}\ }\textbf {\bibinfo {volume} {305}},\ \bibinfo {pages} {1926}
  (\bibinfo {year} {2004})}\BibitemShut {NoStop}%
\bibitem [{\citenamefont {Escaff}\ \emph {et~al.}(2015)\citenamefont {Escaff},
  \citenamefont {Fernandez-Oto}, \citenamefont {Clerc},\ and\ \citenamefont
  {Tlidi}}]{Escaff:PRE:2015}%
  \BibitemOpen
  \bibfield  {author} {\bibinfo {author} {\bibfnamefont {D.}~\bibnamefont
  {Escaff}}, \bibinfo {author} {\bibfnamefont {C.}~\bibnamefont
  {Fernandez-Oto}}, \bibinfo {author} {\bibfnamefont {M.~G.}\ \bibnamefont
  {Clerc}}, \ and\ \bibinfo {author} {\bibfnamefont {M.}~\bibnamefont
  {Tlidi}},\ }\href@noop {} {\bibfield  {journal} {\bibinfo  {journal} {Phys.\
  Rev.\ E}\ }\textbf {\bibinfo {volume} {{91}}},\ \bibinfo {pages} {022924}
  (\bibinfo {year} {{2015}})}\BibitemShut {NoStop}%
\bibitem [{\citenamefont {Meron}(1992)}]{Meron:pr:92}%
  \BibitemOpen
  \bibfield  {author} {\bibinfo {author} {\bibfnamefont {E.}~\bibnamefont
  {Meron}},\ }\href@noop {} {\bibfield  {journal} {\bibinfo  {journal} {Phys.
  Rep.}\ }\textbf {\bibinfo {volume} {218}},\ \bibinfo {pages} {1} (\bibinfo
  {year} {1992})}\BibitemShut {NoStop}%
\bibitem [{\citenamefont {Petrov}\ \emph {et~al.}(1997)\citenamefont {Petrov},
  \citenamefont {Ouyang},\ and\ \citenamefont {Swinney}}]{Swinney:nature:97}%
  \BibitemOpen
  \bibfield  {author} {\bibinfo {author} {\bibfnamefont {V.}~\bibnamefont
  {Petrov}}, \bibinfo {author} {\bibfnamefont {Q.}~\bibnamefont {Ouyang}}, \
  and\ \bibinfo {author} {\bibfnamefont {H.}~\bibnamefont {Swinney}},\
  }\href@noop {} {\bibfield  {journal} {\bibinfo  {journal} {Nature}\ }\textbf
  {\bibinfo {volume} {388}},\ \bibinfo {pages} {655} (\bibinfo {year}
  {1997})}\BibitemShut {NoStop}%
\bibitem [{\citenamefont {Newell}\ \emph {et~al.}(1993)\citenamefont {Newell},
  \citenamefont {Passot},\ and\ \citenamefont {Lega}}]{Newel:arfm:93}%
  \BibitemOpen
  \bibfield  {author} {\bibinfo {author} {\bibfnamefont {A.~C.}\ \bibnamefont
  {Newell}}, \bibinfo {author} {\bibfnamefont {T.}~\bibnamefont {Passot}}, \
  and\ \bibinfo {author} {\bibfnamefont {J.}~\bibnamefont {Lega}},\ }\href@noop
  {} {\bibfield  {journal} {\bibinfo  {journal} {Annu. Rev. Fluid Mech.}\
  }\textbf {\bibinfo {volume} {25}},\ \bibinfo {pages} {399} (\bibinfo {year}
  {1993})}\BibitemShut {NoStop}%
\bibitem [{\citenamefont {Likos}(2001)}]{Likos:PhysRep:2001}%
  \BibitemOpen
  \bibfield  {author} {\bibinfo {author} {\bibfnamefont {C.}~\bibnamefont
  {Likos}},\ }\href@noop {} {\bibfield  {journal} {\bibinfo  {journal} {Phys.
  Rep.}\ }\textbf {\bibinfo {volume} {348}},\ \bibinfo {pages} {267} (\bibinfo
  {year} {2001})}\BibitemShut {NoStop}%
\bibitem [{\citenamefont {Leunissen}\ \emph {et~al.}(2005)\citenamefont
  {Leunissen}, \citenamefont {Christova}, \citenamefont {Hynninen},
  \citenamefont {Royall}, \citenamefont {Campbell}, \citenamefont {Dijkstra},
  \citenamefont {van Roij},\ and\ \citenamefont {van
  Blaaderen}}]{Leunissen:nature:05}%
  \BibitemOpen
  \bibfield  {author} {\bibinfo {author} {\bibfnamefont {M.}~\bibnamefont
  {Leunissen}}, \bibinfo {author} {\bibfnamefont {C.}~\bibnamefont
  {Christova}}, \bibinfo {author} {\bibfnamefont {A.}~\bibnamefont {Hynninen}},
  \bibinfo {author} {\bibfnamefont {C.}~\bibnamefont {Royall}}, \bibinfo
  {author} {\bibfnamefont {A.}~\bibnamefont {Campbell}}, \bibinfo {author}
  {\bibfnamefont {A.}~\bibnamefont {Dijkstra}}, \bibinfo {author}
  {\bibfnamefont {R.}~\bibnamefont {van Roij}}, \ and\ \bibinfo {author}
  {\bibfnamefont {A.}~\bibnamefont {van Blaaderen}},\ }\href@noop {} {\bibfield
   {journal} {\bibinfo  {journal} {Nature}\ }\textbf {\bibinfo {volume}
  {437}},\ \bibinfo {pages} {7056} (\bibinfo {year} {2005})}\BibitemShut
  {NoStop}%
\bibitem [{\citenamefont {Liu}\ \emph {et~al.}(2008)\citenamefont {Liu},
  \citenamefont {Chew},\ and\ \citenamefont {Yu}}]{Liu:pre:08}%
  \BibitemOpen
  \bibfield  {author} {\bibinfo {author} {\bibfnamefont {Y.~H.}\ \bibnamefont
  {Liu}}, \bibinfo {author} {\bibfnamefont {L.~Y.}\ \bibnamefont {Chew}}, \
  and\ \bibinfo {author} {\bibfnamefont {M.~Y.}\ \bibnamefont {Yu}},\
  }\href@noop {} {\bibfield  {journal} {\bibinfo  {journal} {Phys. Rev. E}\
  }\textbf {\bibinfo {volume} {78}},\ \bibinfo {pages} {066405} (\bibinfo
  {year} {2008})}\BibitemShut {NoStop}%
\bibitem [{\citenamefont {Grynberg}\ \emph {et~al.}(1988)\citenamefont
  {Grynberg}, \citenamefont {Le~Bihan}, \citenamefont {Verkerk}, \citenamefont
  {Simoneau}, \citenamefont {Leite}, \citenamefont {Bloch}, \citenamefont
  {LeBoiteux},\ and\ \citenamefont {Ducloy}}]{Grynberg:oc:88}%
  \BibitemOpen
  \bibfield  {author} {\bibinfo {author} {\bibfnamefont {G.}~\bibnamefont
  {Grynberg}}, \bibinfo {author} {\bibfnamefont {E.}~\bibnamefont {Le~Bihan}},
  \bibinfo {author} {\bibfnamefont {P.}~\bibnamefont {Verkerk}}, \bibinfo
  {author} {\bibfnamefont {P.}~\bibnamefont {Simoneau}}, \bibinfo {author}
  {\bibfnamefont {J.}~\bibnamefont {Leite}}, \bibinfo {author} {\bibfnamefont
  {D.}~\bibnamefont {Bloch}}, \bibinfo {author} {\bibfnamefont
  {S.}~\bibnamefont {LeBoiteux}}, \ and\ \bibinfo {author} {\bibfnamefont
  {M.}~\bibnamefont {Ducloy}},\ }\href@noop {} {\bibfield  {journal} {\bibinfo
  {journal} {Opt. Commun.}\ }\textbf {\bibinfo {volume} {67}},\ \bibinfo
  {pages} {363} (\bibinfo {year} {1988})}\BibitemShut {NoStop}%
\bibitem [{\citenamefont {Etrich}\ \emph {et~al.}(1997)\citenamefont {Etrich},
  \citenamefont {Peschel},\ and\ \citenamefont {Lederer}}]{Ulf:pre:97}%
  \BibitemOpen
  \bibfield  {author} {\bibinfo {author} {\bibfnamefont {C.}~\bibnamefont
  {Etrich}}, \bibinfo {author} {\bibfnamefont {U.}~\bibnamefont {Peschel}}, \
  and\ \bibinfo {author} {\bibfnamefont {F.}~\bibnamefont {Lederer}},\ }\href
  {\doibase 10.1103/PhysRevE.56.4803} {\bibfield  {journal} {\bibinfo
  {journal} {Phys. Rev. E}\ }\textbf {\bibinfo {volume} {56}},\ \bibinfo
  {pages} {4803} (\bibinfo {year} {1997})}\BibitemShut {NoStop}%
\bibitem [{\citenamefont {Arecchi}\ \emph {et~al.}(1999)\citenamefont
  {Arecchi}, \citenamefont {Boccaletti},\ and\ \citenamefont
  {Ramazza}}]{Arecchi:pr:99}%
  \BibitemOpen
  \bibfield  {author} {\bibinfo {author} {\bibfnamefont {F.~T.}\ \bibnamefont
  {Arecchi}}, \bibinfo {author} {\bibfnamefont {S.}~\bibnamefont {Boccaletti}},
  \ and\ \bibinfo {author} {\bibfnamefont {P.}~\bibnamefont {Ramazza}},\
  }\href@noop {} {\bibfield  {journal} {\bibinfo  {journal} {Phys. Reports}\
  }\textbf {\bibinfo {volume} {318}},\ \bibinfo {pages} {83} (\bibinfo {year}
  {1999})}\BibitemShut {NoStop}%
\bibitem [{\citenamefont {Lodahl}\ and\ \citenamefont
  {Saffman}(1999)}]{Lodahl:pra:1999}%
  \BibitemOpen
  \bibfield  {author} {\bibinfo {author} {\bibfnamefont {P.}~\bibnamefont
  {Lodahl}}\ and\ \bibinfo {author} {\bibfnamefont {M.}~\bibnamefont
  {Saffman}},\ }\href@noop {} {\bibfield  {journal} {\bibinfo  {journal} {Phys.
  Rev. A}\ }\textbf {\bibinfo {volume} {60}},\ \bibinfo {pages} {3251}
  (\bibinfo {year} {1999})}\BibitemShut {NoStop}%
\bibitem [{\citenamefont {Camara}\ \emph {et~al.}(2015)\citenamefont {Camara},
  \citenamefont {Kaiser}, \citenamefont {Labeyrie}, \citenamefont {Firth},
  \citenamefont {Oppo}, \citenamefont {Robb}, \citenamefont {Arnold},\ and\
  \citenamefont {Ackemann}}]{Camara:PRA:2015}%
  \BibitemOpen
  \bibfield  {author} {\bibinfo {author} {\bibfnamefont {A.}~\bibnamefont
  {Camara}}, \bibinfo {author} {\bibfnamefont {R.}~\bibnamefont {Kaiser}},
  \bibinfo {author} {\bibfnamefont {G.}~\bibnamefont {Labeyrie}}, \bibinfo
  {author} {\bibfnamefont {W.~J.}\ \bibnamefont {Firth}}, \bibinfo {author}
  {\bibfnamefont {G.-L.}\ \bibnamefont {Oppo}}, \bibinfo {author}
  {\bibfnamefont {G.~R.~M.}\ \bibnamefont {Robb}}, \bibinfo {author}
  {\bibfnamefont {A.~S.}\ \bibnamefont {Arnold}}, \ and\ \bibinfo {author}
  {\bibfnamefont {T.}~\bibnamefont {Ackemann}},\ }\href@noop {} {\bibfield
  {journal} {\bibinfo  {journal} {Phys.\ Rev.\ A}\ }\textbf {\bibinfo {volume}
  {92}},\ \bibinfo {pages} {{013820}} (\bibinfo {year} {{2015}})}\BibitemShut
  {NoStop}%
\bibitem [{\citenamefont {Bennink}\ \emph {et~al.}(2002)\citenamefont
  {Bennink}, \citenamefont {Wong}, \citenamefont {Marino}, \citenamefont
  {Aronstein}, \citenamefont {Boyd}, \citenamefont {Stroud}, \citenamefont
  {Lukishova},\ and\ \citenamefont {Gauthier}}]{Bennink:PRL:2002}%
  \BibitemOpen
  \bibfield  {author} {\bibinfo {author} {\bibfnamefont {R.~S.}\ \bibnamefont
  {Bennink}}, \bibinfo {author} {\bibfnamefont {V.}~\bibnamefont {Wong}},
  \bibinfo {author} {\bibfnamefont {A.~M.}\ \bibnamefont {Marino}}, \bibinfo
  {author} {\bibfnamefont {D.~L.}\ \bibnamefont {Aronstein}}, \bibinfo {author}
  {\bibfnamefont {R.~W.}\ \bibnamefont {Boyd}}, \bibinfo {author}
  {\bibfnamefont {C.~R.}\ \bibnamefont {Stroud}}, \bibinfo {author}
  {\bibfnamefont {S.}~\bibnamefont {Lukishova}}, \ and\ \bibinfo {author}
  {\bibfnamefont {D.~J.}\ \bibnamefont {Gauthier}},\ }\href@noop {} {\bibfield
  {journal} {\bibinfo  {journal} {Phys. Rev. Lett.}\ }\textbf {\bibinfo
  {volume} {88}},\ \bibinfo {pages} {113901} (\bibinfo {year}
  {2002})}\BibitemShut {NoStop}%
\bibitem [{\citenamefont {Trillo}\ and\ \citenamefont
  {Torruellas}(2001)}]{Trillo:SS:01}%
  \BibitemOpen
  \bibinfo {editor} {\bibfnamefont {S.}~\bibnamefont {Trillo}}\ and\ \bibinfo
  {editor} {\bibfnamefont {W.}~\bibnamefont {Torruellas}},\ eds.,\ \href@noop
  {} {\emph {\bibinfo {title} {Spatial Solitons}}}\ (\bibinfo  {publisher}
  {Springer},\ \bibinfo {address} {Berlin},\ \bibinfo {year}
  {2001})\BibitemShut {NoStop}%
\bibitem [{\citenamefont {Kivshar}\ and\ \citenamefont
  {Agrawal}(2003)}]{KivsharAgrawal:2003}%
  \BibitemOpen
  \bibfield  {author} {\bibinfo {author} {\bibfnamefont {Y.}~\bibnamefont
  {Kivshar}}\ and\ \bibinfo {author} {\bibfnamefont {G.}~\bibnamefont
  {Agrawal}},\ }\href
  {http://www.scopus.com/inward/record.url?eid=2-s2.0-84902433584&partnerID=40&md5=cdd7bd6e99751ffdd6cae5643e045ac7}
  {\emph {\bibinfo {title} {Optical Solitons: From Fibers to Photonic
  Crystals}}}\ (\bibinfo  {publisher} {Academic Press},\ \bibinfo {year}
  {2003})\BibitemShut {NoStop}%
\bibitem [{\citenamefont {Quiroga-Teixeiro}\ and\ \citenamefont
  {Michinel}(1997)}]{Quiroga:josab:97}%
  \BibitemOpen
  \bibfield  {author} {\bibinfo {author} {\bibfnamefont {M.}~\bibnamefont
  {Quiroga-Teixeiro}}\ and\ \bibinfo {author} {\bibfnamefont {H.}~\bibnamefont
  {Michinel}},\ }\href@noop {} {\bibfield  {journal} {\bibinfo  {journal} {J.
  Opt. Soc. Am. B}\ }\textbf {\bibinfo {volume} {14}},\ \bibinfo {pages} {2004}
  (\bibinfo {year} {1997})}\BibitemShut {NoStop}%
\bibitem [{\citenamefont {Malomed}\ \emph {et~al.}(2002)\citenamefont
  {Malomed}, \citenamefont {Crasovan},\ and\ \citenamefont
  {Mihalache}}]{Malomed:phys_d:02}%
  \BibitemOpen
  \bibfield  {author} {\bibinfo {author} {\bibfnamefont {B.~A.}\ \bibnamefont
  {Malomed}}, \bibinfo {author} {\bibfnamefont {L.-C.}\ \bibnamefont
  {Crasovan}}, \ and\ \bibinfo {author} {\bibfnamefont {D.}~\bibnamefont
  {Mihalache}},\ }\href@noop {} {\bibfield  {journal} {\bibinfo  {journal}
  {Physica D}\ }\textbf {\bibinfo {volume} {161}},\ \bibinfo {pages} {187}
  (\bibinfo {year} {2002})}\BibitemShut {NoStop}%
\bibitem [{\citenamefont {Corney}\ and\ \citenamefont
  {Bang}(2001)}]{Corney:pre:01}%
  \BibitemOpen
  \bibfield  {author} {\bibinfo {author} {\bibfnamefont {J.~F.}\ \bibnamefont
  {Corney}}\ and\ \bibinfo {author} {\bibfnamefont {O.}~\bibnamefont {Bang}},\
  }\href@noop {} {\bibfield  {journal} {\bibinfo  {journal} {Phys. Rev. E}\
  }\textbf {\bibinfo {volume} {64}},\ \bibinfo {pages} {047601} (\bibinfo
  {year} {2001})}\BibitemShut {NoStop}%
\bibitem [{\citenamefont {Couairon}\ and\ \citenamefont
  {Mysyrowicz}(2007)}]{Couairon:pr:441:47}%
  \BibitemOpen
  \bibfield  {author} {\bibinfo {author} {\bibfnamefont {A.}~\bibnamefont
  {Couairon}}\ and\ \bibinfo {author} {\bibfnamefont {A.}~\bibnamefont
  {Mysyrowicz}},\ }\href@noop {} {\bibfield  {journal} {\bibinfo  {journal}
  {Phys. Rep.}\ }\textbf {\bibinfo {volume} {441}},\ \bibinfo {pages} {47}
  (\bibinfo {year} {2007})}\BibitemShut {NoStop}%
\bibitem [{\citenamefont {Agrawal}(2001)}]{Agrawal:NFO:01}%
  \BibitemOpen
  \bibfield  {author} {\bibinfo {author} {\bibfnamefont {G.~P.}\ \bibnamefont
  {Agrawal}},\ }\href@noop {} {\emph {\bibinfo {title} {Nonlinear Fiber
  Optics}}},\ \bibinfo {edition} {3rd}\ ed.\ (\bibinfo  {publisher} {Academic
  Press},\ \bibinfo {address} {San Diego},\ \bibinfo {year} {2001})\BibitemShut
  {NoStop}%
\bibitem [{\citenamefont {Mamaev}\ \emph {et~al.}(1996)\citenamefont {Mamaev},
  \citenamefont {Saffman}, \citenamefont {Anderson},\ and\ \citenamefont
  {Zozulya}}]{Mamaev:PRA:1996}%
  \BibitemOpen
  \bibfield  {author} {\bibinfo {author} {\bibfnamefont {A.~V.}\ \bibnamefont
  {Mamaev}}, \bibinfo {author} {\bibfnamefont {M.}~\bibnamefont {Saffman}},
  \bibinfo {author} {\bibfnamefont {D.~Z.}\ \bibnamefont {Anderson}}, \ and\
  \bibinfo {author} {\bibfnamefont {A.~A.}\ \bibnamefont {Zozulya}},\ }\href
  {\doibase 10.1103/PhysRevA.54.870} {\bibfield  {journal} {\bibinfo  {journal}
  {Phys. Rev. A}\ }\textbf {\bibinfo {volume} {54}},\ \bibinfo {pages} {870}
  (\bibinfo {year} {1996})}\BibitemShut {NoStop}%
\bibitem [{\citenamefont {Saffman}\ \emph {et~al.}(2004)\citenamefont
  {Saffman}, \citenamefont {McCarthy},\ and\ \citenamefont
  {Krolikowski}}]{Saffman:JOB:2004}%
  \BibitemOpen
  \bibfield  {author} {\bibinfo {author} {\bibfnamefont {M.}~\bibnamefont
  {Saffman}}, \bibinfo {author} {\bibfnamefont {G.}~\bibnamefont {McCarthy}}, \
  and\ \bibinfo {author} {\bibfnamefont {W.}~\bibnamefont {Krolikowski}},\
  }\href {\doibase {10.1088/1464-4266/6/5/030}} {\bibfield  {journal} {\bibinfo
   {journal} {J. Opt. B}\ }\textbf {\bibinfo {volume} {6}},\ \bibinfo {pages}
  {S397} (\bibinfo {year} {2004})}\BibitemShut {NoStop}%
\bibitem [{\citenamefont {Meier}\ \emph {et~al.}(2004)\citenamefont {Meier},
  \citenamefont {Stegeman}, \citenamefont {Christodoulides}, \citenamefont
  {Silberberg}, \citenamefont {Morandotti}, \citenamefont {Yang}, \citenamefont
  {Salamo}, \citenamefont {Sorel},\ and\ \citenamefont
  {Aitchison}}]{Meier:PRL:2004}%
  \BibitemOpen
  \bibfield  {author} {\bibinfo {author} {\bibfnamefont {J.}~\bibnamefont
  {Meier}}, \bibinfo {author} {\bibfnamefont {G.~I.}\ \bibnamefont {Stegeman}},
  \bibinfo {author} {\bibfnamefont {D.~N.}\ \bibnamefont {Christodoulides}},
  \bibinfo {author} {\bibfnamefont {Y.}~\bibnamefont {Silberberg}}, \bibinfo
  {author} {\bibfnamefont {R.}~\bibnamefont {Morandotti}}, \bibinfo {author}
  {\bibfnamefont {H.}~\bibnamefont {Yang}}, \bibinfo {author} {\bibfnamefont
  {G.}~\bibnamefont {Salamo}}, \bibinfo {author} {\bibfnamefont
  {M.}~\bibnamefont {Sorel}}, \ and\ \bibinfo {author} {\bibfnamefont {J.~S.}\
  \bibnamefont {Aitchison}},\ }\href {\doibase 10.1103/PhysRevLett.92.163902}
  {\bibfield  {journal} {\bibinfo  {journal} {Phys. Rev. Lett.}\ }\textbf
  {\bibinfo {volume} {92}},\ \bibinfo {pages} {163902} (\bibinfo {year}
  {2004})}\BibitemShut {NoStop}%
\bibitem [{\citenamefont {Henin}\ \emph {et~al.}(2010)\citenamefont {Henin},
  \citenamefont {Petit}, \citenamefont {Kasparian}, \citenamefont {Wolf},
  \citenamefont {Jochmann}, \citenamefont {Kraft}, \citenamefont {Bock},
  \citenamefont {Schramm}, \citenamefont {Sauerbrey}, \citenamefont {Nakaema},
  \citenamefont {Stelmaszczyk}, \citenamefont {Rohwetter}, \citenamefont
  {W{\"o}ste}, \citenamefont {Soulez}, \citenamefont {Mauger}, \citenamefont
  {Berg{\'e}},\ and\ \citenamefont {Skupin}}]{Henin:apb:100:77}%
  \BibitemOpen
  \bibfield  {author} {\bibinfo {author} {\bibfnamefont {S.}~\bibnamefont
  {Henin}}, \bibinfo {author} {\bibfnamefont {Y.}~\bibnamefont {Petit}},
  \bibinfo {author} {\bibfnamefont {J.}~\bibnamefont {Kasparian}}, \bibinfo
  {author} {\bibfnamefont {J.-P.}\ \bibnamefont {Wolf}}, \bibinfo {author}
  {\bibfnamefont {A.}~\bibnamefont {Jochmann}}, \bibinfo {author}
  {\bibfnamefont {S.}~\bibnamefont {Kraft}}, \bibinfo {author} {\bibfnamefont
  {S.}~\bibnamefont {Bock}}, \bibinfo {author} {\bibfnamefont {U.}~\bibnamefont
  {Schramm}}, \bibinfo {author} {\bibfnamefont {R.}~\bibnamefont {Sauerbrey}},
  \bibinfo {author} {\bibfnamefont {W.}~\bibnamefont {Nakaema}}, \bibinfo
  {author} {\bibfnamefont {K.}~\bibnamefont {Stelmaszczyk}}, \bibinfo {author}
  {\bibfnamefont {P.}~\bibnamefont {Rohwetter}}, \bibinfo {author}
  {\bibfnamefont {L.}~\bibnamefont {W{\"o}ste}}, \bibinfo {author}
  {\bibfnamefont {C.-L.}\ \bibnamefont {Soulez}}, \bibinfo {author}
  {\bibfnamefont {S.}~\bibnamefont {Mauger}}, \bibinfo {author} {\bibfnamefont
  {L.}~\bibnamefont {Berg{\'e}}}, \ and\ \bibinfo {author} {\bibfnamefont
  {S.}~\bibnamefont {Skupin}},\ }\href@noop {} {\bibfield  {journal} {\bibinfo
  {journal} {App.\ Phys.\ B}\ }\textbf {\bibinfo {volume} {100}},\ \bibinfo
  {pages} {77} (\bibinfo {year} {2010})}\BibitemShut {NoStop}%
\bibitem [{\citenamefont {Staliunas}\ and\ \citenamefont
  {Sanchez-Morcillo}(2003)}]{OptRes}%
  \BibitemOpen
  \bibfield  {author} {\bibinfo {author} {\bibfnamefont {K.}~\bibnamefont
  {Staliunas}}\ and\ \bibinfo {author} {\bibfnamefont {V.~J.}\ \bibnamefont
  {Sanchez-Morcillo}},\ }\href
  {http://www.scopus.com/inward/record.url?eid=2-s2.0-84902433584&partnerID=40&md5=cdd7bd6e99751ffdd6cae5643e045ac7}
  {\emph {\bibinfo {title} {Transverse Patterns in Nonlinear Optical
  Resonators}}},\ Vol.\ \bibinfo {volume} {183}\ (\bibinfo  {publisher}
  {Springer Berlin Heidelberg},\ \bibinfo {year} {2003})\BibitemShut {NoStop}%
\bibitem [{\citenamefont {Firth}\ and\ \citenamefont
  {Par\'e}(1988)}]{ISI:A1988R274000014}%
  \BibitemOpen
  \bibfield  {author} {\bibinfo {author} {\bibfnamefont {W.~J.}\ \bibnamefont
  {Firth}}\ and\ \bibinfo {author} {\bibfnamefont {C.}~\bibnamefont {Par\'e}},\
  }\href@noop {} {\bibfield  {journal} {\bibinfo  {journal} {{Opt.\ Lett.}}\
  }\textbf {\bibinfo {volume} {{13}}},\ \bibinfo {pages} {{1096}} (\bibinfo
  {year} {{1988}})}\BibitemShut {NoStop}%
\bibitem [{\citenamefont {Geddes}\ \emph {et~al.}(1994)\citenamefont {Geddes},
  \citenamefont {Indik}, \citenamefont {Moloney},\ and\ \citenamefont
  {Firth}}]{ISI:A1994PL62700090}%
  \BibitemOpen
  \bibfield  {author} {\bibinfo {author} {\bibfnamefont {J.~B.}\ \bibnamefont
  {Geddes}}, \bibinfo {author} {\bibfnamefont {R.~A.}\ \bibnamefont {Indik}},
  \bibinfo {author} {\bibfnamefont {J.~V.}\ \bibnamefont {Moloney}}, \ and\
  \bibinfo {author} {\bibfnamefont {W.~J.}\ \bibnamefont {Firth}},\ }\href@noop
  {} {\bibfield  {journal} {\bibinfo  {journal} {{Phys.\ Rev.\ A}}\ }\textbf
  {\bibinfo {volume} {{50}}},\ \bibinfo {pages} {{3471}} (\bibinfo {year}
  {{1994}})}\BibitemShut {NoStop}%
\bibitem [{\citenamefont {Akhmanov}\ \emph {et~al.}(1992)\citenamefont
  {Akhmanov}, \citenamefont {Vorontsov}, \citenamefont {Ivanov}, \citenamefont
  {Larichev},\ and\ \citenamefont {Zheleznykh}}]{ISI:A1992GY19100011}%
  \BibitemOpen
  \bibfield  {author} {\bibinfo {author} {\bibfnamefont {S.~A.}\ \bibnamefont
  {Akhmanov}}, \bibinfo {author} {\bibfnamefont {M.~A.}\ \bibnamefont
  {Vorontsov}}, \bibinfo {author} {\bibfnamefont {V.~Y.}\ \bibnamefont
  {Ivanov}}, \bibinfo {author} {\bibfnamefont {A.~V.}\ \bibnamefont
  {Larichev}}, \ and\ \bibinfo {author} {\bibfnamefont {N.~I.}\ \bibnamefont
  {Zheleznykh}},\ }\href@noop {} {\bibfield  {journal} {\bibinfo  {journal}
  {{J.\ Opt.\ Soc.\ Am.\ B}}\ }\textbf {\bibinfo {volume} {{9}}},\ \bibinfo
  {pages} {{78}} (\bibinfo {year} {{1992}})}\BibitemShut {NoStop}%
\bibitem [{\citenamefont {Likos}\ \emph {et~al.}(2007)\citenamefont {Likos},
  \citenamefont {Mladek}, \citenamefont {Gottwald},\ and\ \citenamefont
  {Kahl}}]{Likos07}%
  \BibitemOpen
  \bibfield  {author} {\bibinfo {author} {\bibfnamefont {C.~N.}\ \bibnamefont
  {Likos}}, \bibinfo {author} {\bibfnamefont {B.~M.}\ \bibnamefont {Mladek}},
  \bibinfo {author} {\bibfnamefont {D.}~\bibnamefont {Gottwald}}, \ and\
  \bibinfo {author} {\bibfnamefont {G.}~\bibnamefont {Kahl}},\ }\href@noop {}
  {\bibfield  {journal} {\bibinfo  {journal} {J.\ Chem.\ Phys.}\ }\textbf
  {\bibinfo {volume} {126}},\ \bibinfo {pages} {224502} (\bibinfo {year}
  {2007})}\BibitemShut {NoStop}%
\bibitem [{\citenamefont {Ultanir}\ \emph {et~al.}(2003)\citenamefont
  {Ultanir}, \citenamefont {Michaelis}, \citenamefont {Lederer},\ and\
  \citenamefont {Stegeman}}]{Ultanir:ol:03}%
  \BibitemOpen
  \bibfield  {author} {\bibinfo {author} {\bibfnamefont {E.~A.}\ \bibnamefont
  {Ultanir}}, \bibinfo {author} {\bibfnamefont {D.}~\bibnamefont {Michaelis}},
  \bibinfo {author} {\bibfnamefont {F.}~\bibnamefont {Lederer}}, \ and\
  \bibinfo {author} {\bibfnamefont {G.~I.}\ \bibnamefont {Stegeman}},\
  }\href@noop {} {\bibfield  {journal} {\bibinfo  {journal} {Opt. Lett.}\
  }\textbf {\bibinfo {volume} {28}},\ \bibinfo {pages} {251} (\bibinfo {year}
  {2003})}\BibitemShut {NoStop}%
\bibitem [{\citenamefont {Dabby}\ and\ \citenamefont
  {Whinnery}(1968)}]{Dabby:apl:13:284}%
  \BibitemOpen
  \bibfield  {author} {\bibinfo {author} {\bibfnamefont {F.~W.}\ \bibnamefont
  {Dabby}}\ and\ \bibinfo {author} {\bibfnamefont {J.~B.}\ \bibnamefont
  {Whinnery}},\ }\href@noop {} {\bibfield  {journal} {\bibinfo  {journal}
  {Appl.\ Phys.\ Lett.}\ }\textbf {\bibinfo {volume} {13}},\ \bibinfo {pages}
  {284} (\bibinfo {year} {1968})}\BibitemShut {NoStop}%
\bibitem [{\citenamefont {Derrien}\ \emph {et~al.}(2000)\citenamefont
  {Derrien}, \citenamefont {Henninot}, \citenamefont {Warenghem},\ and\
  \citenamefont {Abbate}}]{Derrien:JOpt:2000}%
  \BibitemOpen
  \bibfield  {author} {\bibinfo {author} {\bibfnamefont {F.}~\bibnamefont
  {Derrien}}, \bibinfo {author} {\bibfnamefont {J.~F.}\ \bibnamefont
  {Henninot}}, \bibinfo {author} {\bibfnamefont {M.}~\bibnamefont {Warenghem}},
  \ and\ \bibinfo {author} {\bibfnamefont {G.}~\bibnamefont {Abbate}},\
  }\href@noop {} {\bibfield  {journal} {\bibinfo  {journal} {Journal of Optics
  A: Pure and Applied Optics}\ }\textbf {\bibinfo {volume} {2}},\ \bibinfo
  {pages} {332} (\bibinfo {year} {2000})}\BibitemShut {NoStop}%
\bibitem [{\citenamefont {Suter}\ and\ \citenamefont
  {Blasberg}(1993)}]{Suter:1993}%
  \BibitemOpen
  \bibfield  {author} {\bibinfo {author} {\bibfnamefont {D.}~\bibnamefont
  {Suter}}\ and\ \bibinfo {author} {\bibfnamefont {T.}~\bibnamefont
  {Blasberg}},\ }\href@noop {} {\bibfield  {journal} {\bibinfo  {journal}
  {Phys. Rev. A}\ }\textbf {\bibinfo {volume} {48}},\ \bibinfo {pages} {4583}
  (\bibinfo {year} {1993})}\BibitemShut {NoStop}%
\bibitem [{\citenamefont {Skupin}\ \emph {et~al.}(2007)\citenamefont {Skupin},
  \citenamefont {Saffman},\ and\ \citenamefont
  {Kr\'olikowski}}]{Skupin:prl:2007}%
  \BibitemOpen
  \bibfield  {author} {\bibinfo {author} {\bibfnamefont {S.}~\bibnamefont
  {Skupin}}, \bibinfo {author} {\bibfnamefont {M.}~\bibnamefont {Saffman}}, \
  and\ \bibinfo {author} {\bibfnamefont {W.}~\bibnamefont {Kr\'olikowski}},\
  }\href@noop {} {\bibfield  {journal} {\bibinfo  {journal} {Phys. Rev. Lett.}\
  }\textbf {\bibinfo {volume} {98}},\ \bibinfo {pages} {263902} (\bibinfo
  {year} {2007})}\BibitemShut {NoStop}%
\bibitem [{\citenamefont {Conti}\ \emph {et~al.}(2005)\citenamefont {Conti},
  \citenamefont {Ruocco},\ and\ \citenamefont {Trillo}}]{Conti:PRL:2005}%
  \BibitemOpen
  \bibfield  {author} {\bibinfo {author} {\bibfnamefont {C.}~\bibnamefont
  {Conti}}, \bibinfo {author} {\bibfnamefont {G.}~\bibnamefont {Ruocco}}, \
  and\ \bibinfo {author} {\bibfnamefont {S.}~\bibnamefont {Trillo}},\
  }\href@noop {} {\bibfield  {journal} {\bibinfo  {journal} {Phys. Rev. Lett.}\
  }\textbf {\bibinfo {volume} {95}},\ \bibinfo {pages} {183902} (\bibinfo
  {year} {2005})}\BibitemShut {NoStop}%
\bibitem [{\citenamefont {Conti}\ \emph {et~al.}(2006)\citenamefont {Conti},
  \citenamefont {Ghofraniha}, \citenamefont {Ruocco},\ and\ \citenamefont
  {Trillo}}]{Conti:PRL:2006b}%
  \BibitemOpen
  \bibfield  {author} {\bibinfo {author} {\bibfnamefont {C.}~\bibnamefont
  {Conti}}, \bibinfo {author} {\bibfnamefont {N.}~\bibnamefont {Ghofraniha}},
  \bibinfo {author} {\bibfnamefont {G.}~\bibnamefont {Ruocco}}, \ and\ \bibinfo
  {author} {\bibfnamefont {S.}~\bibnamefont {Trillo}},\ }\href@noop {}
  {\bibfield  {journal} {\bibinfo  {journal} {Phys. Rev. Lett.}\ }\textbf
  {\bibinfo {volume} {97}},\ \bibinfo {pages} {123903} (\bibinfo {year}
  {2006})}\BibitemShut {NoStop}%
\bibitem [{sup()}]{suppl}%
  \BibitemOpen
  \href@noop {} {}\bibinfo {note} {See Supplemental Material at [URL] for more
  details on the physical implementation, the variational analysis of solitons,
  the propagation dynamics and movies related to Figs. 3, 4 and an infinite
  system.}\BibitemShut {Stop}%
\bibitem [{\citenamefont {Conti}\ \emph {et~al.}(2003)\citenamefont {Conti},
  \citenamefont {Peccianti},\ and\ \citenamefont {Assanto}}]{Conti:prl:2003}%
  \BibitemOpen
  \bibfield  {author} {\bibinfo {author} {\bibfnamefont {C.}~\bibnamefont
  {Conti}}, \bibinfo {author} {\bibfnamefont {M.}~\bibnamefont {Peccianti}}, \
  and\ \bibinfo {author} {\bibfnamefont {G.}~\bibnamefont {Assanto}},\
  }\href@noop {} {\bibfield  {journal} {\bibinfo  {journal} {Phys. Rev. Lett.}\
  }\textbf {\bibinfo {volume} {91}},\ \bibinfo {pages} {073901} (\bibinfo
  {year} {2003})}\BibitemShut {NoStop}%
\bibitem [{\citenamefont {Ghofraniha}\ \emph {et~al.}(2007)\citenamefont
  {Ghofraniha}, \citenamefont {Conti}, \citenamefont {Ruocco},\ and\
  \citenamefont {Trillo}}]{Ghofraniha:PRL:2007}%
  \BibitemOpen
  \bibfield  {author} {\bibinfo {author} {\bibfnamefont {N.}~\bibnamefont
  {Ghofraniha}}, \bibinfo {author} {\bibfnamefont {C.}~\bibnamefont {Conti}},
  \bibinfo {author} {\bibfnamefont {G.}~\bibnamefont {Ruocco}}, \ and\ \bibinfo
  {author} {\bibfnamefont {S.}~\bibnamefont {Trillo}},\ }\href@noop {}
  {\bibfield  {journal} {\bibinfo  {journal} {Phys. Rev. Lett.}\ }\textbf
  {\bibinfo {volume} {99}},\ \bibinfo {pages} {043903} (\bibinfo {year}
  {2007})}\BibitemShut {NoStop}%
\bibitem [{\citenamefont {Conti}\ \emph {et~al.}(2009)\citenamefont {Conti},
  \citenamefont {Fratalocchi}, \citenamefont {Peccianti}, \citenamefont
  {Ruocco},\ and\ \citenamefont {Trillo}}]{Conti:prl:09}%
  \BibitemOpen
  \bibfield  {author} {\bibinfo {author} {\bibfnamefont {C.}~\bibnamefont
  {Conti}}, \bibinfo {author} {\bibfnamefont {A.}~\bibnamefont {Fratalocchi}},
  \bibinfo {author} {\bibfnamefont {M.}~\bibnamefont {Peccianti}}, \bibinfo
  {author} {\bibfnamefont {G.}~\bibnamefont {Ruocco}}, \ and\ \bibinfo {author}
  {\bibfnamefont {S.}~\bibnamefont {Trillo}},\ }\href@noop {} {\bibfield
  {journal} {\bibinfo  {journal} {Phys. Rev. Lett.}\ }\textbf {\bibinfo
  {volume} {102}},\ \bibinfo {pages} {083902} (\bibinfo {year}
  {2009})}\BibitemShut {NoStop}%
\bibitem [{\citenamefont {Warenghem}\ and\ \citenamefont
  {Henninot}(2006)}]{Warenghem:mclc:06}%
  \BibitemOpen
  \bibfield  {author} {\bibinfo {author} {\bibfnamefont {J.~F.}\ \bibnamefont
  {Warenghem}, \bibfnamefont {M.~Blach}}\ and\ \bibinfo {author} {\bibfnamefont
  {J.~F.}\ \bibnamefont {Henninot}},\ }\href@noop {} {\bibfield  {journal}
  {\bibinfo  {journal} {Mol. Cryst. Liq. Cryst.}\ }\textbf {\bibinfo {volume}
  {454}},\ \bibinfo {pages} {297} (\bibinfo {year} {2006})}\BibitemShut
  {NoStop}%
\bibitem [{\citenamefont {Warenghem}\ \emph {et~al.}(2008)\citenamefont
  {Warenghem}, \citenamefont {Blach},\ and\ \citenamefont
  {Henninot}}]{Warenghem:josab:08}%
  \BibitemOpen
  \bibfield  {author} {\bibinfo {author} {\bibfnamefont {M.}~\bibnamefont
  {Warenghem}}, \bibinfo {author} {\bibfnamefont {J.~F.}\ \bibnamefont
  {Blach}}, \ and\ \bibinfo {author} {\bibfnamefont {J.~F.}\ \bibnamefont
  {Henninot}},\ }\href@noop {} {\bibfield  {journal} {\bibinfo  {journal} {J.
  Opt. Soc. Am. B}\ }\textbf {\bibinfo {volume} {25}},\ \bibinfo {pages} {1882}
  (\bibinfo {year} {2008})}\BibitemShut {NoStop}%
\bibitem [{\citenamefont {Benjamin}\ and\ \citenamefont
  {Feir}(1967)}]{Benjamin:jfm:67}%
  \BibitemOpen
  \bibfield  {author} {\bibinfo {author} {\bibfnamefont {T.~B.}\ \bibnamefont
  {Benjamin}}\ and\ \bibinfo {author} {\bibfnamefont {J.~E.}\ \bibnamefont
  {Feir}},\ }\href@noop {} {\bibfield  {journal} {\bibinfo  {journal} {J. Fluid
  Mech.}\ }\textbf {\bibinfo {volume} {27}},\ \bibinfo {pages} {417} (\bibinfo
  {year} {1967})}\BibitemShut {NoStop}%
\bibitem [{\citenamefont {Esbensen}\ \emph {et~al.}(2011)\citenamefont
  {Esbensen}, \citenamefont {Wlotzka}, \citenamefont {Bache}, \citenamefont
  {Bang},\ and\ \citenamefont {Krolikowski}}]{esbensen:pra:11}%
  \BibitemOpen
  \bibfield  {author} {\bibinfo {author} {\bibfnamefont {B.~K.}\ \bibnamefont
  {Esbensen}}, \bibinfo {author} {\bibfnamefont {A.}~\bibnamefont {Wlotzka}},
  \bibinfo {author} {\bibfnamefont {M.}~\bibnamefont {Bache}}, \bibinfo
  {author} {\bibfnamefont {O.}~\bibnamefont {Bang}}, \ and\ \bibinfo {author}
  {\bibfnamefont {W.}~\bibnamefont {Krolikowski}},\ }\href@noop {} {\bibfield
  {journal} {\bibinfo  {journal} {Phys. Rev. A}\ }\textbf {\bibinfo {volume}
  {84}},\ \bibinfo {pages} {053854} (\bibinfo {year} {2011})}\BibitemShut
  {NoStop}%
\bibitem [{\citenamefont {Bespalov}\ and\ \citenamefont
  {Talanov}(1966)}]{Bespalov:JETP:1966}%
  \BibitemOpen
  \bibfield  {author} {\bibinfo {author} {\bibfnamefont {V.~I.}\ \bibnamefont
  {Bespalov}}\ and\ \bibinfo {author} {\bibfnamefont {V.~I.}\ \bibnamefont
  {Talanov}},\ }\href@noop {} {\bibfield  {journal} {\bibinfo  {journal} {JETP
  Lett.}\ }\textbf {\bibinfo {volume} {3}},\ \bibinfo {pages} {471} (\bibinfo
  {year} {1966})}\BibitemShut {NoStop}%
\bibitem [{\citenamefont {Berg{\'e}}(1998)}]{Berge:pr:303:259}%
  \BibitemOpen
  \bibfield  {author} {\bibinfo {author} {\bibfnamefont {L.}~\bibnamefont
  {Berg{\'e}}},\ }\href@noop {} {\bibfield  {journal} {\bibinfo  {journal}
  {Phys. Rep.}\ }\textbf {\bibinfo {volume} {303}},\ \bibinfo {pages} {259}
  (\bibinfo {year} {1998})}\BibitemShut {NoStop}%
\bibitem [{\citenamefont {Landau}(1941)}]{Landau41}%
  \BibitemOpen
  \bibfield  {author} {\bibinfo {author} {\bibfnamefont {L.}~\bibnamefont
  {Landau}},\ }\href@noop {} {\bibfield  {journal} {\bibinfo  {journal} {Phys.\
  Rev.}\ }\textbf {\bibinfo {volume} {60}},\ \bibinfo {pages} {356} (\bibinfo
  {year} {1941})}\BibitemShut {NoStop}%
\bibitem [{\citenamefont {Feynman}\ and\ \citenamefont
  {Cohen}(1956)}]{Feynman56}%
  \BibitemOpen
  \bibfield  {author} {\bibinfo {author} {\bibfnamefont {R.~P.}\ \bibnamefont
  {Feynman}}\ and\ \bibinfo {author} {\bibfnamefont {M.}~\bibnamefont
  {Cohen}},\ }\href@noop {} {\bibfield  {journal} {\bibinfo  {journal} {Phys.\
  Rev.}\ }\textbf {\bibinfo {volume} {102}},\ \bibinfo {pages} {1189} (\bibinfo
  {year} {1956})}\BibitemShut {NoStop}%
\bibitem [{\citenamefont {Santos}\ \emph {et~al.}(2003)\citenamefont {Santos},
  \citenamefont {Shlyapnikov},\ and\ \citenamefont
  {Lewenstein}}]{Santos:prl:03}%
  \BibitemOpen
  \bibfield  {author} {\bibinfo {author} {\bibfnamefont {L.}~\bibnamefont
  {Santos}}, \bibinfo {author} {\bibfnamefont {G.~V.}\ \bibnamefont
  {Shlyapnikov}}, \ and\ \bibinfo {author} {\bibfnamefont {M.}~\bibnamefont
  {Lewenstein}},\ }\href@noop {} {\bibfield  {journal} {\bibinfo  {journal}
  {Phys. Rev. Lett.}\ }\textbf {\bibinfo {volume} {90}},\ \bibinfo {pages}
  {250403} (\bibinfo {year} {2003})}\BibitemShut {NoStop}%
\bibitem [{\citenamefont {Henkel}\ \emph {et~al.}(2010)\citenamefont {Henkel},
  \citenamefont {Nath},\ and\ \citenamefont {Pohl}}]{Henkel:PRL:2010}%
  \BibitemOpen
  \bibfield  {author} {\bibinfo {author} {\bibfnamefont {N.}~\bibnamefont
  {Henkel}}, \bibinfo {author} {\bibfnamefont {R.}~\bibnamefont {Nath}}, \ and\
  \bibinfo {author} {\bibfnamefont {T.}~\bibnamefont {Pohl}},\ }\href@noop {}
  {\bibfield  {journal} {\bibinfo  {journal} {Phys. Rev. Lett.}\ }\textbf
  {\bibinfo {volume} {104}},\ \bibinfo {pages} {195302} (\bibinfo {year}
  {2010})}\BibitemShut {NoStop}%
\bibitem [{\citenamefont {Schneider}\ and\ \citenamefont
  {Enz}(1971)}]{Schneider71}%
  \BibitemOpen
  \bibfield  {author} {\bibinfo {author} {\bibfnamefont {T.}~\bibnamefont
  {Schneider}}\ and\ \bibinfo {author} {\bibfnamefont {C.~P.}\ \bibnamefont
  {Enz}},\ }\href@noop {} {\bibfield  {journal} {\bibinfo  {journal} {Phys.\
  Rev.\ Lett.}\ }\textbf {\bibinfo {volume} {27}},\ \bibinfo {pages} {1186}
  (\bibinfo {year} {1971})}\BibitemShut {NoStop}%
\bibitem [{\citenamefont {Astrakharchik}\ \emph {et~al.}(2007)\citenamefont
  {Astrakharchik}, \citenamefont {Boronat}, \citenamefont {Kurbakov},\ and\
  \citenamefont {Lozovik}}]{Astrakharchik07}%
  \BibitemOpen
  \bibfield  {author} {\bibinfo {author} {\bibfnamefont {G.~E.}\ \bibnamefont
  {Astrakharchik}}, \bibinfo {author} {\bibfnamefont {J.}~\bibnamefont
  {Boronat}}, \bibinfo {author} {\bibfnamefont {I.~L.}\ \bibnamefont
  {Kurbakov}}, \ and\ \bibinfo {author} {\bibfnamefont {Y.~E.}\ \bibnamefont
  {Lozovik}},\ }\href@noop {} {\bibfield  {journal} {\bibinfo  {journal}
  {Phys.\ Rev.\ Lett.}\ }\textbf {\bibinfo {volume} {98}},\ \bibinfo {pages}
  {060405} (\bibinfo {year} {2007})}\BibitemShut {NoStop}%
\bibitem [{\citenamefont {Cinti}\ \emph
  {et~al.}(2014{\natexlab{a}})\citenamefont {Cinti}, \citenamefont {Macri},
  \citenamefont {Lechner}, \citenamefont {Pupillo},\ and\ \citenamefont
  {Pohl}}]{Cinti14}%
  \BibitemOpen
  \bibfield  {author} {\bibinfo {author} {\bibfnamefont {F.}~\bibnamefont
  {Cinti}}, \bibinfo {author} {\bibfnamefont {T.}~\bibnamefont {Macri}},
  \bibinfo {author} {\bibfnamefont {W.}~\bibnamefont {Lechner}}, \bibinfo
  {author} {\bibfnamefont {G.}~\bibnamefont {Pupillo}}, \ and\ \bibinfo
  {author} {\bibfnamefont {T.}~\bibnamefont {Pohl}},\ }\href@noop {} {\bibfield
   {journal} {\bibinfo  {journal} {Nature\ Comm.}\ }\textbf {\bibinfo {volume}
  {5}},\ \bibinfo {pages} {3235} (\bibinfo {year}
  {2014}{\natexlab{a}})}\BibitemShut {NoStop}%
\bibitem [{\citenamefont {Macri}\ \emph {et~al.}(2013)\citenamefont {Macri},
  \citenamefont {Maucher}, \citenamefont {Cinti},\ and\ \citenamefont
  {Pohl}}]{Macri13}%
  \BibitemOpen
  \bibfield  {author} {\bibinfo {author} {\bibfnamefont {T.}~\bibnamefont
  {Macri}}, \bibinfo {author} {\bibfnamefont {F.}~\bibnamefont {Maucher}},
  \bibinfo {author} {\bibfnamefont {F.}~\bibnamefont {Cinti}}, \ and\ \bibinfo
  {author} {\bibfnamefont {T.}~\bibnamefont {Pohl}},\ }\href@noop {} {\bibfield
   {journal} {\bibinfo  {journal} {Phys.\ Rev.\ A}\ }\textbf {\bibinfo {volume}
  {87}},\ \bibinfo {pages} {061602} (\bibinfo {year} {2013})}\BibitemShut
  {NoStop}%
\bibitem [{\citenamefont {Cinti}\ \emph
  {et~al.}(2014{\natexlab{b}})\citenamefont {Cinti}, \citenamefont
  {Boninsegni},\ and\ \citenamefont {Pohl}}]{Cinti14b}%
  \BibitemOpen
  \bibfield  {author} {\bibinfo {author} {\bibfnamefont {F.}~\bibnamefont
  {Cinti}}, \bibinfo {author} {\bibfnamefont {M.}~\bibnamefont {Boninsegni}}, \
  and\ \bibinfo {author} {\bibfnamefont {T.}~\bibnamefont {Pohl}},\ }\href@noop
  {} {\bibfield  {journal} {\bibinfo  {journal} {New.\ J.\ Phys.}\ }\textbf
  {\bibinfo {volume} {16}},\ \bibinfo {pages} {033038} (\bibinfo {year}
  {2014}{\natexlab{b}})}\BibitemShut {NoStop}%
\end{thebibliography}%

%%%%%%%%%% Merge with supplemental materials %%%%%%%%%%
% \pagebreak
\widetext
\newpage
\begin{center}
\textbf{\large Supplemental Materials: \\Self-organization of light in optical media with competing nonlinearities}
\end{center}
%%%%%%%%%% Merge with supplemental materials %%%%%%%%%%
%%%%%%%%%% Prefix a "S" to all equations, figures, tables and reset the counter %%%%%%%%%%
\setcounter{equation}{0}
\setcounter{figure}{0}
\setcounter{table}{0}
\setcounter{page}{1}
\makeatletter
\renewcommand{\theequation}{S\arabic{equation}}
\renewcommand{\thefigure}{S\arabic{figure}}
\renewcommand{\bibnumfmt}[1]{[S#1]}
\renewcommand{\citenumfont}[1]{S#1}
%%%%%%%%%% Prefix a "S" to all equations, figures, tables and reset the counter %%%%%%%%%%

\section{Physical realization}
\begin{figure}[t!]
\includegraphics[width=0.9\columnwidth]{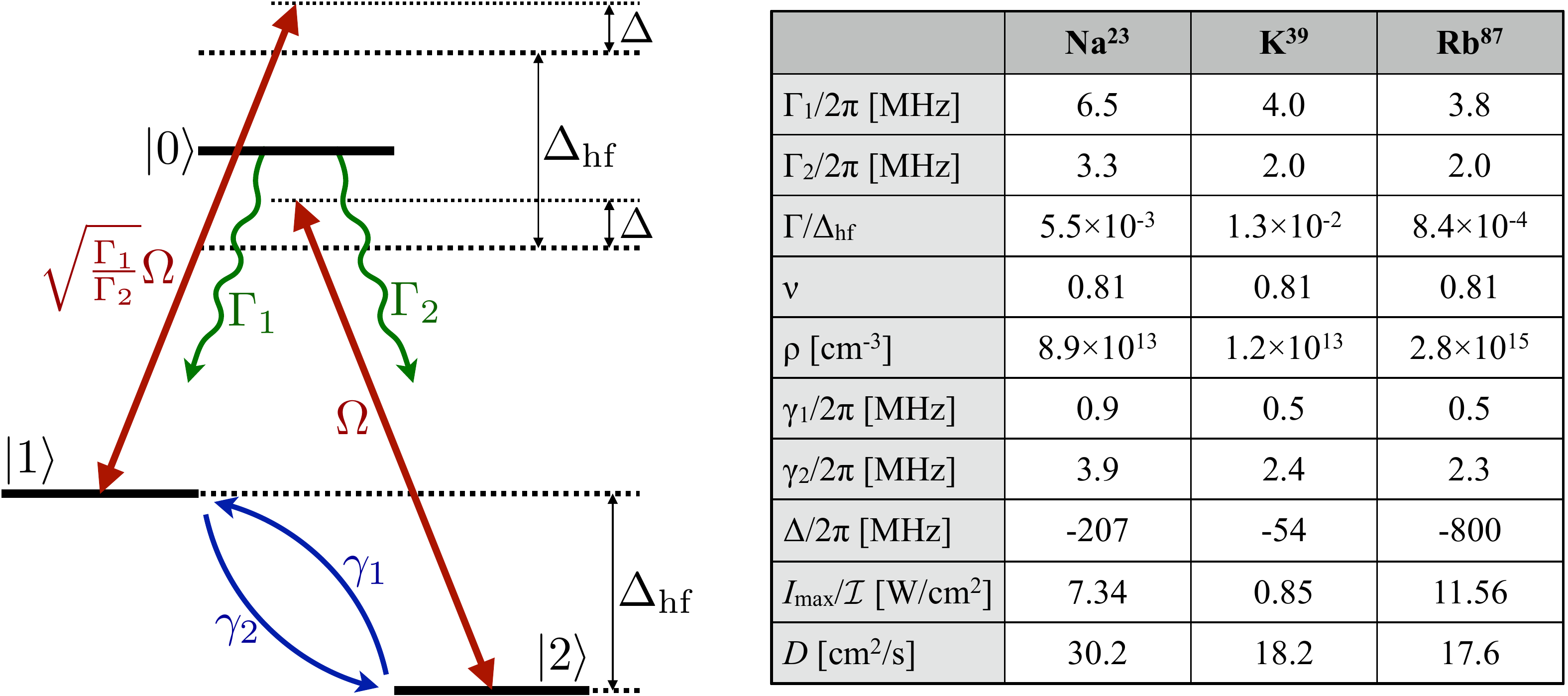}
\caption{Schematic level diagram as described in the text. The table provides the physical parameters corresponding to the dimensionless parameters ($\alpha=1.4$, $\sigma=0.7$, $\beta=0.08$, $\ell=5.3$) used for the numerical simulations described in the main text. The given diffusion constants are chosen to provide a transverse length scale of $\sigma_2=10\mu$m.
\label{figS1}}
\end{figure}

We consider a gas of alkaline atoms with two hyperfine ground state manifolds that are split by an energy $\hbar\Delta_{\rm hf}$. The incident laser beam couples both of them to an excited state manifold. The gas temperature is chosen such that the associated Doppler broadening may completely cover the hyperfine splittings of the excited manifold while leaving the much larger ground state splitting fully resolvable. Under these conditions the atoms can be described by three effective levels: two hyperfine ground states, $|1\rangle=|nS_{1/2}F\rangle$ and $|2\rangle=|nS_{1/2}F^\prime\rangle$, and one excited fine structure state $|0\rangle=|nP_J\rangle$ (see Fig.~\ref{figS1}). Generally, the two transitions feature different dipole matrix elements $\mu_{1(2)}$ resulting in distinct decay rates $\Gamma_{1(2)}$ and Rabi frequencies $\Omega_{1(2)}=\mu_{1(2)}\mathcal{E}/\hbar$ for a single laser field with a slowly varying electric field amplitude $\mathcal{E}$.
In addition, we consider incoherent driving between the states $|1\rangle$ and $|2\rangle$ with rates $\gamma_{1(2)}$, which can be realized either through collisions, broad band microwave coupling or optical pumping via some auxiliary excited states. 
Accounting for diffusive atomic motion with a diffusion constant $D$ the corresponding optical Bloch equations for the population and coherence densities of the gas read
\begin{subequations}\label{eq1}
\begin{eqnarray}
\label{eq1a}
\partial_t\rho_{11}&=&\frac{i}{2}\left(\Omega_1\rho_{01}-\Omega_1^*\rho_{10}\right)+\Gamma_1\rho_{00}-\gamma_2\rho_{11}+\gamma_1\rho_{22}+D\nabla^2\rho_{11},\\
\label{eq1b}
\partial_t\rho_{22}&=&\frac{i}{2}\left(\Omega_2\rho_{02}-\Omega_2^*\rho_{20}\right)+\Gamma_2\rho_{00}+\gamma_2\rho_{11}-\gamma_1\rho_{22}+D\nabla^2\rho_{22},\\
\label{eq1c}
\partial_t\rho_{10}&=&i\frac{\Omega_1}{2}\left(\rho_{00}-\rho_{11}\right)-i\frac{\Omega_2}{2}\rho_{12}+i\Delta_1\rho_{10}-\frac{\Gamma}{2}\rho_{10},\\
\label{eq1d}
\partial_t\rho_{20}&=&i\frac{\Omega_2}{2}\left(\rho_{00}-\rho_{22}\right)-i\frac{\Omega_1}{2}\rho_{21}+i\Delta_2\rho_{20}-\frac{\Gamma}{2}\rho_{20},\\
\label{eq1e}
\partial_t\rho_{12}&=&i\frac{\Omega_1}{2}\rho_{02}-i\frac{\Omega_2^*}{2}\rho_{10}+i(\Delta_1-\Delta_2)\rho_{12}.
\end{eqnarray}
\end{subequations}
Here, we have neglected diffusion of the coherence densities, since the large laser detunings $|\Delta_i|\gg\Gamma$ lead to a negligibly small range, $\sim |\Delta_i|^{-1}$, of the corresponding nonlocality which is much smaller than the corresponding length scales, $\sim \Gamma_i^{-1}$, arising from population diffusion (see below).

In order to determine the steady state of Eqs.~(\ref{eq1}) we first calculate the stationary coherences up to third order in the light field amplitude
\begin{subequations}\label{eq2}
\begin{eqnarray}
\label{eq2a}
\rho_{12}&=&-\frac{\Omega_1}{2\Delta_{\rm hf}}\rho_{02}+\frac{\Omega_2^*}{2\Delta_{\rm hf}}\rho_{10},\\
\label{eq2b}
\rho_{10}&=&-\frac{\Omega_1}{(2\Delta_1+i\Gamma)}\left(\rho_{00}-\rho_{11}\right)-\frac{\Omega_1|\Omega_2|^2}{2\Delta_{\rm hf}(2\Delta_1+i\Gamma)^2}\left(\rho_{00}-\rho_{11}\right)+\frac{\Omega_1|\Omega_2|^2}{2\Delta_{\rm hf}(2\Delta_1+i\Gamma)(2\Delta_2-i\Gamma)}\left(\rho_{00}-\rho_{22}\right),\\
\label{eq2c}
\rho_{20}&=&-\frac{\Omega_2}{2\Delta_2+i\Gamma}\left(\rho_{00}-\rho_{22}\right)+\frac{|\Omega_1|^2\Omega_2}{2\Delta_{\rm hf}(2\Delta_2+i\Gamma)^2}\left(\rho_{00}-\rho_{22}\right)-\frac{|\Omega_1|^2\Omega_2}{2\Delta_{\rm hf}(2\Delta_1-i\Gamma)(2\Delta_2+i\Gamma)}\left(\rho_{00}-\rho_{11}\right),
\end{eqnarray}
\end{subequations}
where $\Delta_1-\Delta_2=\Delta_{\rm hf}\gg\Omega_{1(2)}$. As we shall see below, the first term in Eqs.~(\ref{eq2b}) and (\ref{eq2c}) gives rise to a linear optical response and a nonlocal nonlinearity while the second and third terms provide a local nonlinearity. 

Substituting, Eqs.~(\ref{eq2}) into Eqs.~(\ref{eq1a}) and (\ref{eq1b}) gives a closed expression for the dynamics of the hyperfine populations 
\begin{subequations}\label{eq3}
\begin{eqnarray}
\partial_t\rho_{11}&=&\tilde{\Gamma}_1\left(\rho_{00}-\rho_{11}\right)+\Gamma_1\rho_{00}-\gamma_2\rho_{11}+\gamma_1\rho_{22}+D\nabla^2\rho_{11},\\
\partial_t\rho_{22}&=&\tilde{\Gamma}_2\left(\rho_{00}-\rho_{22}\right)+\Gamma_2\rho_{00}+\gamma_2\rho_{11}-\gamma_1\rho_{22}+D\nabla^2\rho_{22},
\end{eqnarray}
\end{subequations}
where
\begin{equation}\label{eq4}
\tilde{\Gamma}_i=\frac{\Gamma}{\Gamma^2+4\Delta_i^2}|\Omega_i|^2\;.
\end{equation}
Again we perform an expansion in $\Omega_i/\Delta_i$ to calculate the populations up to $2^{\rm nd}$ order in the field amplitude $\mathcal{E}$. The $0^{\rm th}$ order is particularly simple 
\begin{equation}\label{eq5}
\rho_{11}^{(0)}=\frac{\gamma_1}{\gamma}\rho\;,\quad\rho_{22}^{(0)}=\frac{\gamma_2}{\gamma}\rho.
\end{equation}
where $\gamma=\gamma_1+\gamma_2$ is the total hyperfine pumping rate and $\rho=\rho_{00}+\rho_{11}+\rho_{22}$ is the atomic density. Substituting this result into Eqs.~(\ref{eq3}) yields the following steady state relation for the $2^{\rm nd}$ order 
\begin{subequations}\label{eq6}
\begin{eqnarray}\label{eq6a}
0&=&-\tilde{\Gamma}_1\frac{\gamma_1}{\gamma}\rho-(\Gamma_1+\gamma_2)\rho_{11}^{(2)}-(\Gamma_1-\gamma_1)\rho_{22}^{(2)}+D\nabla^2\rho_{11}^{(2)},\\
\label{eq6b}
0&=&-\tilde{\Gamma}_2\frac{\gamma_2}{\gamma}\rho-(\Gamma_2-\gamma_2)\rho_{11}^{(2)}-(\Gamma_2+\gamma_1)\rho_{22}^{(2)}+D\nabla^2\rho_{22}^{(2)}.
\end{eqnarray}
\end{subequations}
These equations can be conveniently solved in Fourier space which, after some simple algebra, yields a sum of modified Bessel functions, $K_0$, 
\begin{subequations}\label{eq7}
\begin{eqnarray}
\label{eq7a}
\rho_{11}^{(2)}({\bf r})&=&-\frac{\sigma_1^{-2}\rho}{\gamma\Gamma(\Gamma-\gamma)}\int{\rm d}{\bf r}^{\prime}K_0\left(|{\bf r}-{\bf r}^\prime|/\sigma_1\right)\left[\kappa_1\gamma_1\tilde{\Gamma}_1({\bf r}^\prime)+\kappa_1\gamma_2\tilde{\Gamma}_2({\bf r}^\prime)\right]\nonumber\\
&&-\frac{\sigma_2^{-2}\rho}{\gamma^2(\Gamma-\gamma)}\int{\rm d}{\bf r}^{\prime}K_0\left(|{\bf r}-{\bf r}^\prime|/\sigma_2\right)\left[\kappa_2\gamma_1\tilde{\Gamma}_1({\bf r}^\prime)-\kappa_1\gamma_2\tilde{\Gamma}_2({\bf r}^\prime)\right],\\
\label{eq7b}
\rho_{22}^{(2)}({\bf r})&=&-\frac{\sigma_1^{-2}\rho}{\gamma\Gamma(\Gamma-\gamma)}\int{\rm d}{\bf r}^{\prime}K_0\left(|{\bf r}-{\bf r}^\prime|/\sigma_1\right)\left[\kappa_2\gamma_1\tilde{\Gamma}_1({\bf r}^\prime)+\kappa_2\gamma_2\tilde{\Gamma}_2({\bf r}^\prime)\right]\nonumber\\
&&+\frac{\sigma_2^{-2}\rho}{\gamma^2(\Gamma-\gamma)}\int{\rm d}{\bf r}^{\prime}K_0\left(|{\bf r}-{\bf r}^\prime|/\sigma_2\right)\left[\kappa_2\gamma_1\tilde{\Gamma}_1({\bf r}^\prime)-\kappa_1\gamma_2\tilde{\Gamma}_2({\bf r}^\prime)\right],
\end{eqnarray}
\end{subequations}
where $\kappa_i=\Gamma_i-\gamma_i$. The two emerging length scales of the nonlocal response are given by $\sigma_1^2=D/\Gamma$ and $\sigma_2^2=D/\gamma$. Having obtained the laser-driven steady state of the atomic gas, we can now formulate the paraxial wave equation for the electric field amplitude of the propagating light. It is most conveniently written in terms of the Rabi frequency $\Omega\equiv\Omega_2=\sqrt{\Gamma_2/\Gamma_1}\Omega_1$,
\begin{equation}\label{eq8}
\left(-\frac{i}{2k}\nabla^2+\partial_z\right)\Omega=-i\frac{3\pi}{k^2}\left(\sqrt{\Gamma_1\Gamma_2}\rho_{10}+\Gamma_2\rho_{20}\right).
\end{equation}
Substituting Eqs.~(\ref{eq2b}) and (\ref{eq2c}) together with Eqs.~(\ref{eq5}) and (\ref{eq7}) yields a closed equation for the nonlinear field propagation. To simplify the resulting expression we choose the laser frequency in between the two frequencies of the hyperfine transitions, and only retain leading order terms in the detuning $\Delta=(\Delta_1+\Delta_2)/2$ from this central frequency (see Fig.\ref{figS1}). For further simplification we restrict the following discussion to the $D_1$ transition of alkaline atoms with a nuclear spin of $3/2$, such as Na$^{23}$, K$^{39}$ and Rb$^{87}$, for which $\Gamma_1/\Gamma_2=2$. We note that both assumptions are not strictly necessary, and generalizations to larger values of $\Delta$ as well as other elements and transitions are straightforward using Eqs.~(\ref{eq2b}), (\ref{eq2c}), (\ref{eq5}) (\ref{eq7}) and (\ref{eq8}). Scaling the Rabi frequency by its maximum input value $\Omega_{\rm max}=\max_{\bf r} |\Omega_{{\bf r},z=0}|$ allows us
  to write for the dimensionless field amplitude $\varphi=\Omega/\Omega_{\rm max}$
\begin{eqnarray}\label{eq9}
\left(\frac{1}{2k}\nabla^2+i\partial_z\right)\varphi&=&-i\frac{3\pi}{k^2}\frac{\Gamma}{\gamma\Delta_{\rm hf}^2}\left[\gamma_1\Gamma_1+\gamma_2\Gamma_2\right]\rho\varphi-\varepsilon\frac{3\pi}{k^2}\frac{4\Gamma}{3\Delta_{\rm hf}}\frac{\Delta}{\Delta_{\rm hf}}\rho\varphi|\varphi|^2\nonumber\\
&&-\varepsilon\frac{3\pi}{k^2}\frac{\Gamma^2\sigma_1^{-2}\rho}{\Delta_{\rm hf}(\Gamma-\gamma)}\int{\rm d}{\bf r}^{\prime}K_0\left(|{\bf r}-{\bf r}^\prime|/\sigma_1\right)(2-\nu)[2/3+\sigma^2(\nu-1)]|\varphi({\bf r}^\prime)|^2\varphi\nonumber\\
&&+\varepsilon\frac{3\pi}{k^2}\frac{\Gamma^2\sigma_2^{-2}\rho}{\Delta_{\rm hf}(\Gamma-\gamma)}\int{\rm d}{\bf r}^{\prime}K_0\left(|{\bf r}-{\bf r}^\prime|/\sigma_2\right)\frac{4\nu+3\sigma^2\nu-3\nu^2\sigma^2-2}{3\sigma^2}|\varphi({\bf r}^\prime)|^2\varphi,
\end{eqnarray}
where $\nu=\gamma_2/\gamma$, $\sigma=\sigma_1/\sigma_2=\sqrt{\gamma/\Gamma}$ and $\epsilon=\Omega_{\rm max}^2/\Delta_{\rm hf}^2$. Finally, we scale the transverse coordinate by $\sigma_2$ and $z$ by $2k\sigma_2^2$ to obtain Eqs.~(1) and (2) of the main text, i.e.,
\begin{eqnarray}\label{eq10}
\left(\nabla^2+i\partial_z\right)\psi&=&-i\ell^{-1}\psi+\beta|\psi|^2\psi-\alpha\int{\rm d}{\bf r}^{\prime}K_0\left(|{\bf r}-{\bf r}^\prime|/\sigma\right)|\psi({\bf r}^\prime)|^2\psi+\int{\rm d}{\bf r}^{\prime}K_0\left(|{\bf r}-{\bf r}^\prime|\right)|\psi({\bf r}^\prime)|^2\psi,
\end{eqnarray}
with the dimensionless parameters
\begin{equation}\label{eq11}
\ell^{-1}=\left(\frac{\sigma_2}{\lambda}\right)^2\frac{\Gamma^2}{\Delta_{\rm hf}^2}\left[2-\nu\right]\bar{\rho},
\end{equation}
\begin{equation}\label{eq12}
\alpha=\frac{(2-\nu)[2+3\sigma^2(\nu-1)]}{\left[4\nu-3\sigma^2\nu(\nu-1)-2\right]},
\end{equation}
\begin{equation}\label{eq13}
\beta=-\frac{4\sigma^{2}(1-\sigma^2)}{4\nu+3\sigma^2\nu-3\nu^2\sigma^2-2}\frac{\Delta}{\Delta_{\rm hf}},
\end{equation}
and the maximum input value of the dimensionless field amplitude
\begin{equation}\label{eq14}
\mathcal{I}=\varepsilon\frac{\Delta_{\rm hf}}{\Gamma}\frac{\ell^{-1}}{(2-\nu)}\frac{\sigma^{-2}}{(1-\sigma^2)}\left[4\nu+3\sigma^2\nu-3\nu^2\sigma^2-2\right],
\end{equation}
where $\psi=\mathcal{I}\varphi$.

From these expressions we can determine the required physical parameters for any chosen set of the dimensionless quantities $\alpha$, $\sigma$, $\beta$, $\ell$ and $\mathcal{I}$, introduced in the main text. Equation~(\ref{eq12}) fixes $\nu$ for given values of $\alpha$ and $\sigma$, which yields explicit values for the hyperfine pump rates $\gamma_{1}$ and $\gamma_{2}$ from $\nu=\gamma_2/\gamma$ and $\sigma=\sqrt{\gamma/\Gamma}$. Having determined $\nu$, the detuning $\Delta$ is readily obtained from Eq.~(\ref{eq13}) for a given $\beta$. The dimensionless absorption length $\ell$ fixes the required density for a given transverse length scale $\sigma_2$ via Eq.~(\ref{eq11}). Finally, we can rewrite Eq.~(\ref{eq14}) to express the peak input intensity 
\begin{equation}\label{eq15}
I_{\rm max}=X\frac{\Delta_{\rm hf}}{\Gamma}I_{\rm sat}\mathcal{I},
\end{equation}
in terms of the dimensionless intensity $\mathcal{I}$, with
\begin{equation}\label{eq16}
X=\frac{6\ell(2-\nu)\sigma^{2}(1-\sigma^2)}{4\nu+3\sigma^2\nu-3\nu^2\sigma^2-2}
\end{equation}
and the saturation intensity $I_{\rm sat}=\pi h\Gamma c/(3\lambda^3)$ of the $D_1$-transition. Here, $h$ denotes the Planck constant and $c$ the speed of light. A precise tuning of the diffusion constant $D$, is not necessary for observing the phenomena described in the main text, as it merely determines the transverse unit length $\sigma_2=\sqrt{D/\gamma}$. Typical values of all of these parameters corresponding to the conditions used in the dynamical simulations presented in the main text are listed in the table of Fig.~\ref{figS1} for different elements.

\section{Soliton existence}
The soliton solution is approximated by a Gaussian variational ansatz
\begin{equation}\label{eq17}
\psi_{\rm sol}({\bf r})=\frac{\sqrt{\mathcal{P}}}{\sqrt{2\pi s^2}}{\rm e}^{-r^2/4s^2}{\rm e}^{i\mu z}
\end{equation}
with an RMS width $s$ and total power $\mathcal{P}=\int{\rm d}{\bf r}|\psi_{\rm sol}({\bf r})|^2$. To analyze its stability we consider the corresponding Hamiltonian density $H[\psi_{\rm sol}]$ [Eq.~(5) of the main text] relative to that of an unstructured plane wave with $|\psi_{\rm pw}|^2=\mathcal{P}/V=\mathcal{I}$. The involved integrals can be evaluated analytically and amount to 
\begin{equation}\label{eq18}
H[\psi_{\rm sol}]=\mathcal{I}\left[\frac{1}{2s^2}+\frac{\beta\mathcal{P}}{8\pi s^2}+\frac{\mathcal{P}}{4}{\rm Ei}(1,s^2){\rm e}^{s^2}-\frac{\alpha\mathcal{P}}{4}{\rm Ei}(1,s^2/\sigma^2){\rm e}^{s^2/\sigma^2}\right]
\end{equation}
for the variational soliton solution.
We find that the variational solitons are in excellent agreement with numerical ones obtained from imaginary propagation.
However, we have to keep in mind that $H[\psi_{\rm sol}]$ is always (slightly) larger than the Hamiltonian value of the exact soliton. The exponential integral ${\rm Ei}(1,x)$ can be written as
\begin{equation}
 {\rm Ei}(1,x)=x \int_0^1\int_0^1 e^{-x \zeta_1\zeta_2}d\zeta_1 d\zeta_2 -\gamma_e-{\rm ln}(x),
\end{equation}
where $\gamma_e$ denotes the Euler-Mascheroni constant. 
On the other hand, for the plane wave solution we find
\begin{equation}\label{eq19}
H[\psi_{\rm pw}]=\frac{\mathcal{I}^2}{2}\left[\beta+2\pi-2\pi\alpha\sigma^2\right].
\end{equation}
For an infinitely extended system ($V\rightarrow\infty$), we can neglect the kinetic energy term in Eq.~(\ref{eq18}), since $\mathcal{P}=\mathcal{I}V\rightarrow\infty$. For the same reason, the absolute value $|H[\psi_{\rm sol}]|\propto \mathcal{P}$ of the energy density exceeds that of the plane wave for any value of $s$. Hence, it suffices to require $H[\psi_{\rm sol}]\ge0$ in order to prevent a solitonic groundstate. The corresponding $\beta_{\rm cr}$ is readily obtained by rewriting Eq.~(\ref{eq18}) as
\begin{equation}\label{eq20}
H[\psi_{\rm sol}]=\frac{\mathcal{P}\mathcal{I}}{8\pi s^2}\left[{\beta+f(s)}\right]\;.
\end{equation}
Since the function
\begin{equation}\label{eq21}
f(s)=2\pi s^2{\rm Ei}(1,s^2){\rm e}^{s^2}-2\pi s^2\alpha{\rm Ei}(1,s^2/\sigma^2){\rm e}^{s^2/\sigma^2}
\end{equation}
always has a negative minimum if $\alpha>1$ and $0<\sigma<1$, the condition 
\begin{equation}\label{eq22}
\beta>\beta_{\rm cr}=-\min_s f(s)
\end{equation}
assures that $H[\psi_{\rm sol}]>H[\psi_{\rm pw}]$ and, hence, prevents the formation of a soliton. We have used Eqs.~(\ref{eq21}) and (\ref{eq22}) to determine $\beta_{\rm cr}$ and calculate the intensity map $\mathcal{I}_{\rm MI}(\alpha,\sigma; \beta_{\rm cr})$ shown in Fig.~1(a) of the main text. Because Eq.~(\ref{eq18}) systematically overestimates the Hamiltonian value of the exact soliton solution (see remark above), this procedure estimates $\beta_{\rm cr}$ from below.

Along the phase boundary $\alpha=\sigma^{-4}$ we can use the asymptotic expression ${\rm Ei}(1,x)={\rm e}^{-x}/x$ ($x\gg1$) to simplify Eq.~(\ref{eq21}) to $f(s)=2\pi-2\pi \sigma^{-2}$. This yields
\begin{equation}\label{eq23}
\beta_{\rm cr}=2\pi(\sigma^{-2}-1)\;,
\end{equation}
which coincides with the exact expression derived in the main text.

\section{Conservative propagation dynamics}
\begin{figure}[t!]
\includegraphics[width=.85\columnwidth]{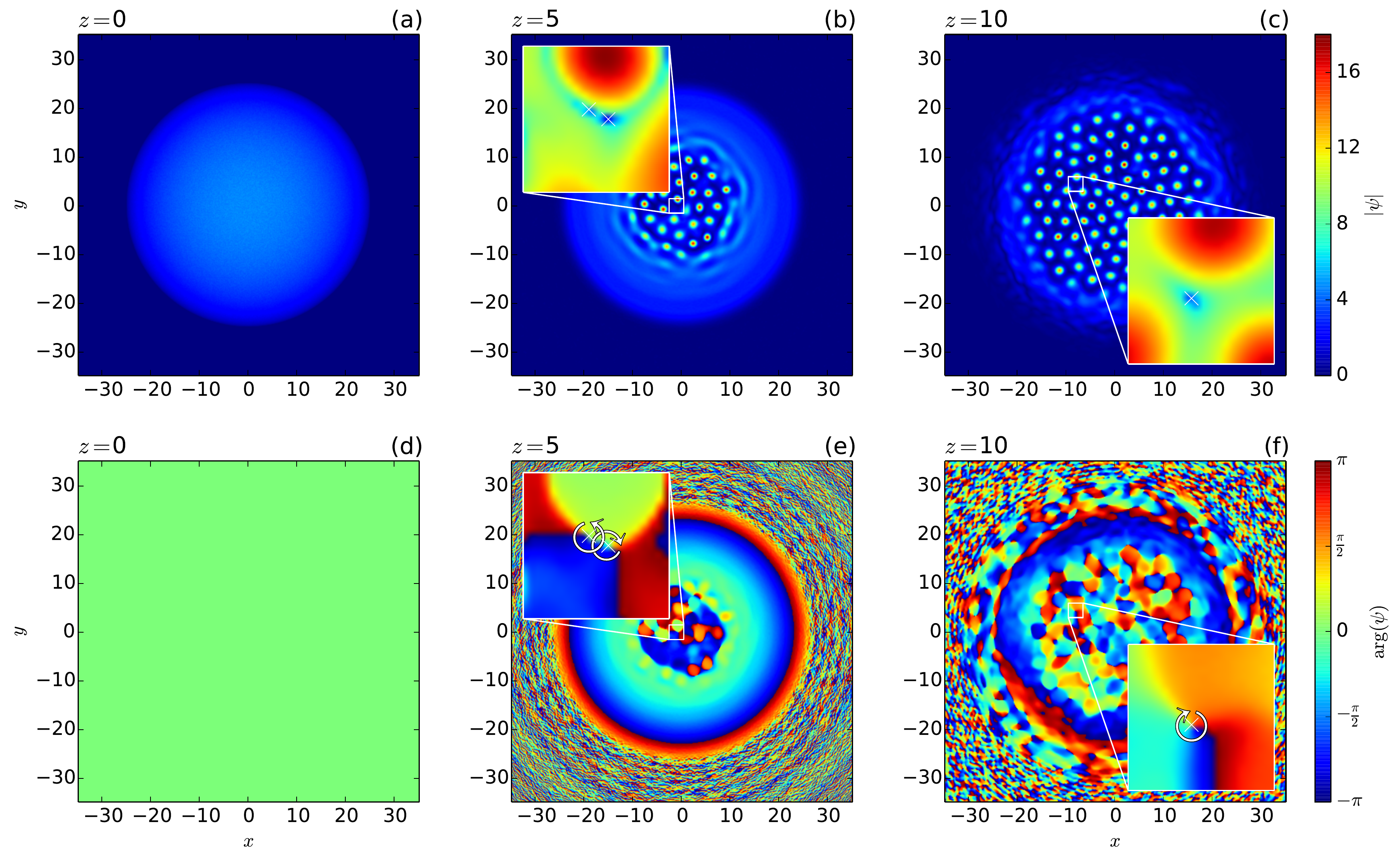}
\caption{Guided light propagation for $\alpha=1.4$, $\sigma=0.7$, $\beta=0.08$, $U({\bf r})=(r/4)^2$, $\ell^{-1}=0$ and intensity $\mathcal{I}=20>\mathcal{I}_{\rm MI}$ at
different propagation distances [(c,f) corresponds to Fig.~3(b,c) in the main text]. The initial beam profile (a) has a flat phase (d). When pattern formation kicks in (b,c), the phase evolution become more complex and phase singularities appear (e,f). The insets of panels (e,f) provide a closer look at such singularities and indicate their topological phase charge, while the insets of panels (e,f) show the corresponding intensity minima at each phase singularity, highlighted through a logarithmic color coding. 
\label{figS2}}
\end{figure}
Figure~\ref{figS2} provides more details on the propagation dynamics depicted in Fig.~3(b,c) of the main text. The intensity snapshots shown Figs.~\ref{figS2}(b,c) (see also corresponding movie) show that the initial beam profile settles into a hexagonal lattice structure close to the groundstate of the system, i.e. the state featuring the minimal Hamiltonian. Note, however, that for the considered conservative dynamics ($\ell^{-1}=0$) the total Hamiltonian
\begin{equation}
H= \int \left|\nabla\psi({\bf r})\right|^{2}{\rm d}^2r -
\iint  R(\mathbf{r}-\mathbf{r}^\prime)\left|\psi_{\rm st}({\bf r})\right|^{2}|\psi_{\rm st}(\mathbf{r}^\prime)|^{2}{\rm d}^2r^\prime {\rm d}^2r
\end{equation}
is a conserved quantity and considerably larger than ground state Hamiltonian of the ordered state. Yet, pattern formation is still possible by decreasing the Hamiltonian 
\begin{equation}
H_{|\psi|}= \int \left(\nabla|\psi({\bf r})|\right)^{2}{\rm d}^2r -
\iint  R(\mathbf{r}-\mathbf{r}^\prime)\left|\psi({\bf r})\right|^{2}|\psi(\mathbf{r}^\prime)|^{2}{\rm d}^2r^\prime {\rm d}^2r
\end{equation}
associated with the light intensity $|\psi|$. While $H_{|\psi|}$ eventually becomes close to the ground state value, the excess Hamiltonian is taken up by simultaneously forming phase inhomogeneities, i.e. it is dissipated into the Hamiltonian
\begin{equation}
H_{\varphi}= \int \left|\psi({\bf r})\right|^{2}\left[\nabla\varphi({\bf r})\right]^{2}{\rm d}^2r
\end{equation}
associated with phase $\varphi=\arg(\psi)$ of the light field, such that $H_{|\psi|}+H_{\varphi}=H={\rm const.}$. The emergence and evolution of these phase inhomogeneities is depicted in Figs.~\ref{figS2}(e,f) (see also corresponding movie and Fig.~3(c)  of the main text). Most of the surplus Hamiltonian $H_{\varphi}$ is carried by phase singularities that eventually settle in between the intensity peaks, as exemplarily shown in the insets of Figs.~\ref{figS2}(e,f). Conserving the total topological phase charge, such singularities form in pairs of opposite charge $\pm 1$, 
as indicated in Fig.~\ref{figS2}(e). The corresponding zero-intensity spots expected at the singularity are indicated in the insets of Figs.~\ref{figS2}(b,c). Such scenario also holds for the infinite system, as confirmed by numerical simulations employing an initially constant intensity and periodic boundary conditions (see {\tt movie\_infinite.mp4}).

\end{document}